\documentclass{aa}  

\usepackage{graphicx}
\usepackage{txfonts}
\usepackage[colorlinks = true,
            linkcolor = blue,
            urlcolor  = black,
            citecolor = cyan,
            anchorcolor = black]{hyperref}

\newcommand {\ha} {H$\alpha$}

\newcommand{\hi}{H\,\textsc{i}}

\newcommand {\kms} {\,{\rm km\,s}^{-1}}

\newcommand {\um} {\,{\mu\rm m}}
\newcommand {\kpc} {\,{\rm kpc}}
\newcommand {\Mpc} {\,{\rm Mpc}}

\newcommand {\msun}{\,{\rm M}_\odot}

\def\apjl{ApJ}%
\def\apjs{ApJS}%
\def\ao{Appl.~Opt.}%
\def\aap{A\&A}%
\defcitealias{Posti+19}{PFM19}

\begin{document} 

\renewcommand{\figureautorefname}{Figure\!}
\title{Massive disc galaxies in cosmological hydrodynamical simulations are too dark matter-dominated}
\titlerunning{Massive discs in cosmological simulations}
\authorrunning{A. Marasco et al.}

   \author{A. Marasco\inst{1},
          L. Posti\inst{2},
          K. Oman\inst{3},
          B. Famaey\inst{2},
          G. Cresci\inst{1}
          and
          F. Fraternali\inst{4}
          }

    \institute{INAF - Osservatorio Astrofisico di Arcetri, Largo E. Fermi 5, 50127, Firenze, Italy\\
                  \email{antonino.marasco@inaf.it}
         \and
                 Universit\'e de Strasbourg, CNRS UMR 7550, Observatoire astronomique de Strasbourg, 11 rue de l'Universit\'e, 67000 Strasbourg, France
        \and
                Institute for Computational Cosmology, Department of Physics, Durham University, South Road, Durham DH1 3LE, UK
        \and
            Kapteyn Astronomical Institute, University of Groningen, Postbus 800, 9700 AV Groningen, The Netherlands
             }

   \date{Received ; accepted}

 
\abstract
{We investigate the disc-halo connection in massive ($M_\star>5\times10^{10}\msun$) disc galaxies from the cosmological hydrodynamical simulations EAGLE and IllustrisTNG, and compare it with that inferred from the study of \hi\ rotation curves in nearby massive spirals from the Spitzer Photometry and Accurate Rotation Curves (SPARC) dataset.
We find that discrepancies between the simulated and observed discs arise both on global and on local scales.
Globally, the simulated discs inhabit halos that are a factor $\sim4$ (in EAGLE) and $\sim2$ (in IllustrisTNG) more massive than those derived from the rotation curve analysis of the observed dataset. We also use synthetic rotation curves of the simulated discs to demonstrate that the recovery of the halo masses from rotation curves are not systematically biased.
We find that the simulations predict dark-matter dominated systems with stellar-to-total enclosed mass ratios that are a factor of $1.5-2$ smaller than real galaxies at all radii.
This is an alternative manifestation of the `failed feedback problem', since it indicates that simulated halos hosting massive discs have been too inefficient at converting their baryons into stars, possibly due to an overly efficient stellar and/or AGN feedback implementation.}

\maketitle
%

\section{Introduction}
In the standard $\Lambda$ cold dark matter ($\Lambda$CDM) framework, galaxies form via the cooling and gravitational collapse of baryonic matter within the potential wells provided by the dark matter halos \citep[e.g.][]{WhiteRees78}.
Assuming a universal baryonic-to-dark matter fraction, $f_{\rm b}\equiv\Omega_b/\Omega_c\simeq0.188$ \citep{Planck+18} one should expect that, on average, halos of mass $M_{\rm halo}$ host gas reservoirs with masses of $f_{\rm b} M_{\rm halo}$ out of which galaxies can form.
However, the efficiency of the baryons-to-stars conversion process, $f_\star=M_\star/(f_{\rm b} M_{\rm halo})$, along with the morphological, kinematic and chemical properties of the resulting system, depends on the complex interplay between the various physical processes that orchestrate galaxy evolution, and cannot easily be predicted a priori.
Independent estimates of the so-called galaxy-halo connection at different masses, different epochs and for galaxies of different morphological types are required to provide constraints on the whole theoretical framework of galaxy formation.

One of the key ingredients of the galaxy-halo connection is the relation between $M_\star$ and $M_{\rm halo}$ (or, equivalently, between $M_\star$ and $f_\star$), the stellar-to-halo mass relation \citep[SHMR, see][for a recent review]{WechslerTinker18}.
This relation is commonly probed via a semi-empirical technique known as abundance matching (AM), which relates central galaxies to halos by matching the observed galaxy stellar mass function to the theoretical halo mass function, under the assumption that stellar mass increases monotonically with the mass of the host halo \citep{ValeOstriker04, Behroozi+10, Moster+13, Kravtsov+18}.
Taken together, different AM studies build up a coherent picture where $f_\star$ peaks at $\sim20\%$ in $L_\star$ galaxies, and rapidly decreases at lower and higher masses. 
Such global inefficiency of baryon-to-star conversion is interpreted as evidence for `negative' feedback from star formation itself (for $M_{\rm halo}\!\lesssim10^{12}\!\msun$) and Active Galactic Nuclei (AGN) activity (for $M_{\rm halo}\!\gtrsim10^{12}\!\msun$).

Observationally, the SHMR can be probed via different techniques such as galaxy-galaxy weak lensing \citep{Mandelbaum+06, Leauthaud+12}, satellite kinematics \citep{Bosch+04, More+11, Bosch+19}, internal galaxy dynamics \citep{Persic+96, Cappellari+13, Read+17} or a combination of these \citep{Dutton+10}.
While generally confirming the scenario predicted by AM techniques, some of these studies have signalled a bimodality in the SHMR for the most luminous late- and early-type systems, with the former systematically occupying halos with $M_{\rm halo}\!<\!10^{13}\msun$, and the latter being preferentially located in groups and clusters with $M_{\rm halo}\!>\!10^{13}\msun$.
However, the paucity of spirals at $M_\star>10^{11}\msun$ makes precise measurements challenging, and it is unclear whether the observed bimodality arises naturally from the shape and scatter of the SHMR \citep{Moster+19} or is symptomatic of different star formation efficiencies associated to different galaxy types \citep{Mandelbaum+16}.

Recently, \citet[][hereafter \citetalias{Posti+19}]{Posti+19} have determined the SHMR in a sample of nearby isolated disc galaxies from the Spitzer Photometry and Accurate Rotation Curves \citep[SPARC,][]{Lelli+16} dataset via the mass decomposition of their \hi\ rotation curves.
Their results show the existence of a monotonic SHMR for discs spanning more than four orders of magnitude in $M_\star$, with the most massive spirals inhabiting `light' dark matter halos and having $f_\star$ close to unity, in striking contrast with predictions from AM methods.
The existence of such a monotonic SHMR is intimately connected to the monotonicity of the relations between the stellar masses, sizes and rotational velocities of discs \citep{Posti+19b}, and is evidence for the presence of different pathways for the formation of early- and late-type galaxies.
This result is not per se incompatible with AM, assuming that the high-mass end of the galaxy stellar mass function is dominated by early-type systems, but outlines the existence of a class of galaxies for which feedback has failed at quenching the star formation efficiency \citep[the `failed feedback problem',][]{Posti+19b}.
Nonetheless, while this discrepancy was noted on global scales, the main culprit, i.e. feedback, acts on the scales of galactic discs.
This leads us to ask whether the detailed structure of discs is also affected by this phenomenon; in other words, whether the local dynamical structure of real massive spirals behaves as expected from current state-of-the-art models.

In this work we compare these observational results with the predictions from two state-of-the-art cosmological hydrodynamical simulation suites, EAGLE \citep{Schaye+15} and IllustrisTNG \citep{Pillepich+18}.
While the parameters of these simulations are tuned to reproduce a number of observables at $z\!=\!0$, including the galaxy stellar mass function, the detailed connection between galaxies and their hosting halos is not forced `by hand' but follows from the complex physics of galaxy formation, which is treated self-consistently.
These models are adequate to resolve the morphology and internal dynamics for several tens of massive spirals at $z\!=\!0$, which makes them the best possible tools to investigate the connection between galaxy type and the SHMR.

\section{Simulated and observed galaxy samples}\label{method}
\begin{figure*}
\begin{center}
\includegraphics[width=1.0\textwidth]{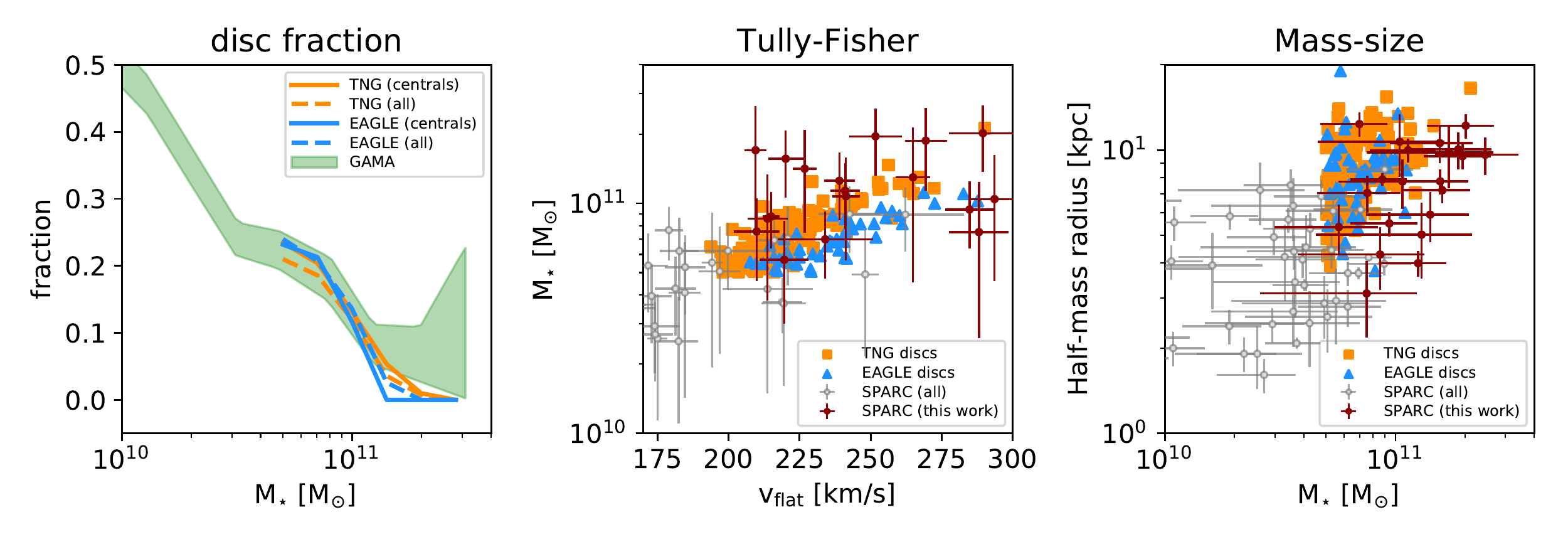}
\caption{\emph{Left panel:} fraction of disc galaxies as a function of their stellar mass in EAGLE (blue lines) and IllustrisTNG (orange lines), compared to that measured in the GAMA survey by \citet[][shaded green area]{Moffett+16}. Solid lines show centrals only; dashed lines include also satellites.
\emph{Central panel:} stellar Tully-Fisher relation for our subsample of simulated (central) discs with $M_\star\!>\!5\times10^{10}\msun$ in EAGLE (blue triangles) and IllustrisTNG (orange squares), compared with the population of nearby spirals from the SPARC dataset (circles with error bars). Galaxies from SPARC are shown as circles with error-bars. Filled red circles are used for the sub-sample of massive discs studied in this work. \emph{Right panel:} stellar mass-size relation for the same systems.} 
\label{fig:scaling_relations}
\end{center}
\end{figure*}

We focus our analysis on isolated, regularly-rotating disc galaxies with stellar masses $M_{\star}$ larger than $5\times10^{10}\msun$.
As shown by \citetalias{Posti+19}, in this mass range the derived $f_\star$ of discs diverges significantly from the behaviour predicted by AM methods.
Our observed sample consists therefore of the $21$ massive discs selected by \citetalias{Posti+19} in this mass range.
We note that the original SPARC sample includes additional $11$ massive galaxies which have been excluded in the study of \citetalias{Posti+19}, either because of their low inclination ($2$) or because the rotation curve modelling led to a poor inference on $M_\star$ or $M_{\rm halo}$ ($9$).
The latter case comprises also $3$ edge-on spirals for which the rotation curve in the inner regions suffers from projection effects.

We build our simulated galaxy sample using two suites of very well-known, publicly available cosmological hydrodynamical simulations of galaxy formation in the $\Lambda$CDM framework: EAGLE and IllustrisTNG.
Both simulation suites follow self-consistently the formation and evolution of galaxies and of their environments, and include treatments for star formation, stellar evolution, black-hole accretion, feedback from supernovae and AGN, primordial and metal-line gas cooling and, in the case of IllustrisTNG, the amplification and evolution of seed magnetic fields.
The parameters of both models are calibrated to output a `realistic' population of galaxies at $z\!=\!0$ in terms of their number densities, sizes, central black-hole masses and star formation rates.
Differences between the predictions of the two models are most often caused by: differences in the treatment of the `sub-grid' physics (e.g. stellar and AGN feedback implementation); differences in the accuracy with which the model calibration succeeds in reproducing the observed calibrators; the inclusion of magnetic field physics in IllustrisTNG (absent in EAGLE); and the use of different solvers for the (magneto\nobreakdash-)hydrodynamical equations\footnote{EAGLE uses a modified version of the SPH code {\sc gadget-2} \citep{gadget2} while IllustrisTNG uses the moving-mesh code {\sc arepo} \citep{Springel10}.}.
Further details on these simulations can be found in \citet{Schaye+15,Crain+15} and \citet{Pillepich+18}.

The runs which we consider here are Ref-L0100N1504 (in EAGLE) and TNG100-1 (in IllustrisTNG).
The former (latter) considers a cubic volume with side length of $100\Mpc$ ($111\Mpc$) and uses dark matter particles with a mass of $9.7$ ($7.4$)$\times10^6\msun$, gas particles (cells) with initial masses $1.8$ ($1.4$)$\times10^6\msun$, and a gravitational softening length of $0.70$ ($0.74$)$\kpc$.
Thus, both runs are adequate to resolve the morphology and the kinematics of hundreds of galaxies in the mass range of interest.

Using the EAGLE and IllustrisTNG galaxy catalogues from the public releases of \citet{McAlpine+16} and \citet{Nelson+19}, we select all central subhalos at $z\!=\!0 $ in our stellar mass range of interest\footnote{A `central' galaxy is the most massive subhalo of a friend-of-friends group. In each subhalo, $M_\star$ is computed within a sphere of $30\kpc$ of radius centred on the minimum of the gravitational potential.}.
We focus our study on centrals given that the SPARC spirals do not show clear signs of major interactions, neither they lie in the proximity of more massive systems.
In order to extract a subsample of regularly rotating disc galaxies, we use two morpho-kinematical estimators: the ratio between stellar rotational velocity and velocity dispersion ($\mathcal{R}_\star$), and the stellar disc fraction ($\mathcal{F}_\star$).
The former is given by the ratio between the mass-weighted median rotational speed for stars orbiting within the galactic plane and their velocity dispersion perpendicular to it, while the latter is based on the fraction of non-counter-rotating stars within $R<30\kpc$ \citep{Thob+19}.
While both estimators were already available in the EAGLE catalogues, only $\mathcal{F}_\star$ was pre-computed for IllustrisTNG, so we determined $\mathcal{R}_\star$ for our subsample using the procedure of \citet{Thob+19}.

We label as `discs' those systems having $\mathcal{R}_\star>1.7$ and $\mathcal{F}_\star>0.7$.
These thresholds ensure that resulting fraction of disc galaxies decreases as a function of $M_\star$, following a trend which is compatible with observations from the GAMA survey \citep{Driver+11} derived by \citet{Moffett+16}, as we show in the first panel of Fig.\,\ref{fig:scaling_relations}.
We stress that the comparison is purely qualitative, as the morphological classification used in GAMA is based on visual inspection of optical and near-infrared images.
Yet, both EAGLE and IllustrisTNG galaxies seem to be in agreement with the observed trend.
As an additional step, we visually inspected the morphology of the simulated discs using the (synthetic) edge-on and face-on composite optical images available from the two simulation databases, and discarded those (few) galaxies which either appeared to be strongly warped or showed visible signs of recent interactions with companions.
These selections resulted in a final sample of $46$ systems for EAGLE and $130$ systems for IllustrisTNG.
Virtually all galaxies in this sample occupy halos with $12<\log(M_{\rm halo}/\msun)<12.7$, consistent with Local Group-like environments.
We stress, and discuss below, that our results do not depend on the adopted thresholds for $\mathcal{R}_\star$ and $\mathcal{F}_\star$.
Tables listing the main properties of the galaxies studied in this work, along with examples of synthetic optical images are presented in Appendix \ref{app:supplementary}.

In the central panel of Fig.\,\ref{fig:scaling_relations} we compare the stellar Tully-Fisher relation \citep[TFR,][]{TullyFisher77} for our sample of simulated galaxies and for the SPARC sample.
The definition of $v_{\rm flat}$ - the velocity at which the rotation curve flattens - used for the simulated discs is based on their circular velocity profile derived assuming spherical symmetry, $v_{\rm c}\!=\!\sqrt{GM(<R)/R}$, which we compute for each galaxy in our sample, along with its decomposition into the separate contributions of stars, gas (in the case of IllustrisTNG, including `wind' particles), and dark matter.
To ensure the best possible comparison with the data, we truncate our $v_{\rm c}$ profiles at the `\hi\ radius' as determined from the \hi\ mass-size relation of \citet{Lelli+16}\footnote{We note that discs in EAGLE and IllustrisTNG reproduces well this relation \citep{Bahe+16,Diemer+19}.}, under the assumption that the simulated discs have an \hi\ content analogous to that of SPARC galaxies with similar stellar mass.
This stratagem allows us to bypass the various pitfalls that arise when dealing with rotation curves derived directly from `simulated' \hi\ data \citep[e.g.][see also Section \ref{discussion}]{Oman+19}, and is based on the ansatz that \hi\ rotation curves in real galaxies are excellent proxies for $v_{\rm c}$.
We have verified visually that the vast majority of $v_{\rm c}$ profiles flatten out in the outer disc regions (see Appendix \ref{app:supplementary}) and that rotational speeds extracted in the proximity of $R_{\rm HI}$ are good proxies for $v_{\rm flat}$.
We therefore set $v_{\rm flat}$ to the mean velocity measured in the interval between $R_{\rm HI}$ and $3\kpc$ inward of this radius.

In general, there is a very good agreement between the simulated and the observed data, with the former producing a very narrow sequence in the $v_{\rm flat}-M_{\star}$ plane passing in between the SPARC data points.
However, important differences appear at $M_\star\!\gtrsim\!10^{11}\msun$, where most simulated discs have large ($>250\kms$) $v_{\rm flat}$ while observed spirals show a wider distribution of rotational speeds, with a mean shifted towards lower velocities.
In the right panel of Fig.\,\ref{fig:scaling_relations} we compare the size-$M_\star$ relation for the SPARC and the simulated galaxies, under the assumption that the $3.6\um$ effective radii, used for the observed sample, are good proxies for the half-$M_\star$ radii, used for the simulated sample.
Also in this case the agreement is good, but less accurate, with the simulated galaxies occupying preferentially the upper tail of the observed size distribution at fixed $M_\star$.

In summary, overall good agreement exists between simulated and real discs in terms of global scaling relations, although the former have slightly larger sizes and rotational speeds than the latter, especially at the high-M$_\star$ end.
As we show below, these small differences become more evident when investigating the disc-halo connection.

\section{Results}\label{results}
\subsection{Global disc-halo connection}\label{ssec:SHMR_global}
\citetalias{Posti+19} determined the stellar and dark matter content of SPARC galaxies via the analysis of their \hi\ rotation curves and of their $3.6\um$ \emph{Spitzer} photometry.
Their Bayesian approach led them to infer unimodal, well defined posteriors on the mass-to-light ratio and on $M_{\rm halo}$ for $137$ galaxies.
In the simulations, we have the luxury of knowing precisely the stellar and dark matter content of our galaxies, which opens the possibility to two complementary approaches: we can either use the stellar and halo masses reported in the catalogues, or carry out rotation curve decompositions using the same $v_{\rm c}$ profiles discussed in Section \ref{method}.
We present the results derived with the former approach below, while in Appendix \ref{app:mass_dec} we demonstrate that they do not change if we use the latter, more observationally-oriented method.
This is a confirmation of the validity and robustness of the methodology of \citetalias{Posti+19}: mass-decomposition of \hi\ rotation curves is a powerful tool to determine halo masses in a $\Lambda$CDM universe, at least in the mass regime studied here.

\begin{figure*}
\begin{center}
\includegraphics[width=0.7\textwidth]{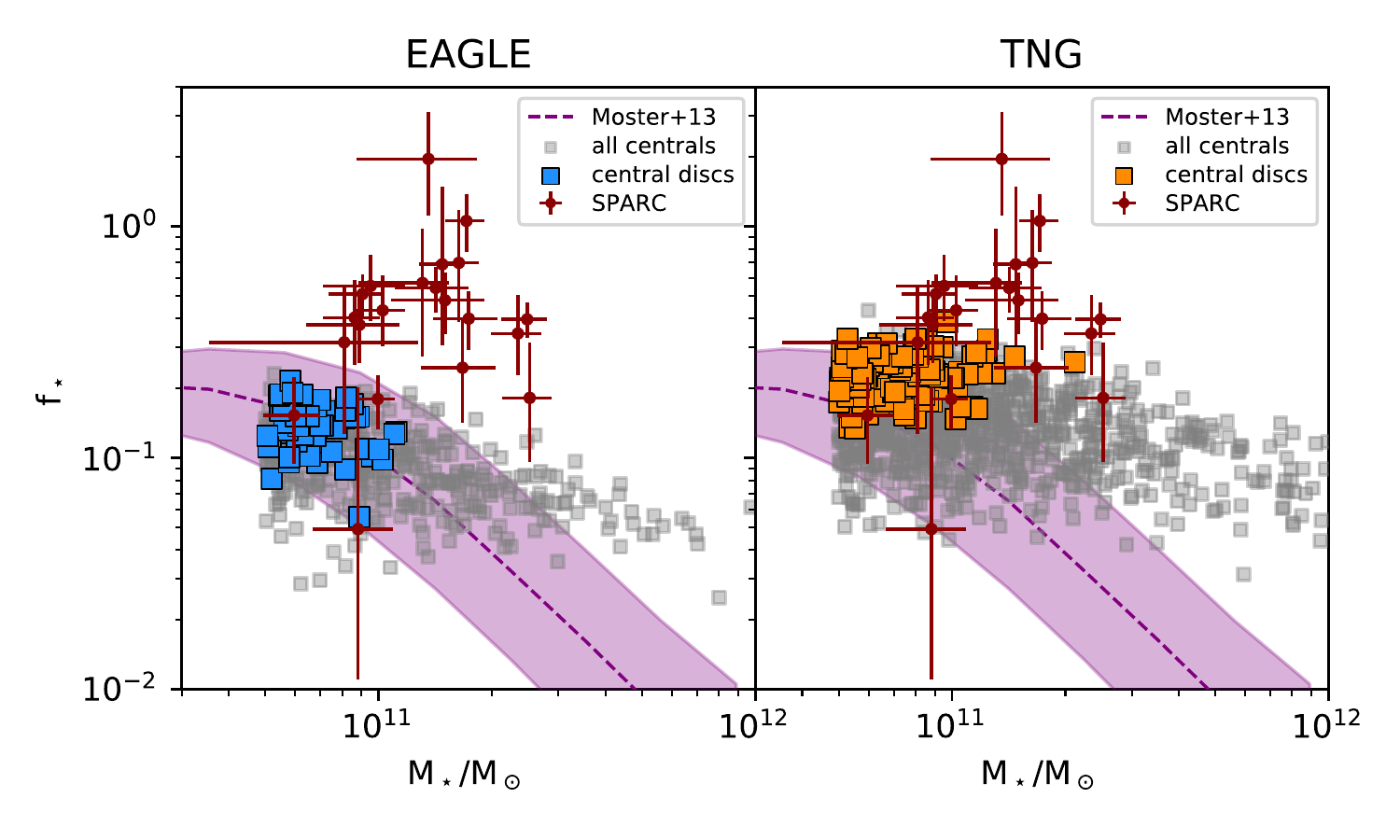}
\caption{Stellar fraction as a function of the stellar mass for simulated centrals in EAGLE (left panel) and IllustrisTNG (right panel) with $M_\star>5\times10^{10}\msun$, compared to nearby spirals from the SPARC dataset (red circles with error-bar). Coloured symbols are used for our subsample of simulated discs. The purple shaded region shows the AM relation and related scatter from \citet{Moster+13}.} 
\label{fig:shmr}
\end{center}
\end{figure*}

In Fig.\,\ref{fig:shmr} we show the relation between $f_\star$ and $M_\star$ for the whole population of massive centrals in EAGLE and IllustrisTNG, for the subsample of simulated discs and for the SPARC sample, along with the prediction from the AM method of \citet{Moster+13}.
In general, the SPARC massive discs have higher $f_\star$ compared to the simulated galaxies, which is expected given that spheroids dominate in the high-mass regime.
On average, at fixed $M_\star$, IllustrisTNG centrals have a higher $f_\star$ than EAGLE centrals due to an overall higher (by a factor of $\sim2$) normalization of the stellar mass function in this mass range, which implies that more numerous (i.e., less massive) halos are populated by galaxies of that $M_\star$.
This factor $\sim2$ higher normalization, combined with the $1.4$ difference in the simulated volume, explains why we find a larger abundance of massive discs in IllustrisTNG with respect to EAGLE ($130$ vs $46$).

Moving our focus to disc galaxies, we notice that EAGLE and IllustrisTNG discs systematically occupy the high end of the $f_\star$ distribution, having on average $35\%$ ($0.13$ dex) higher $f_\star$ than the typical simulated galaxy with the same stellar mass.
However, this is not enough to match the exceptionally high $f_\star$ of the observed spirals, which sit well above predictions from both AM methods and hydrodynamical simulations.
The mismatch is expected given the offset in the TFR at high $M_\star$ (Fig.\,\ref{fig:scaling_relations}), but is also due to an additional offset in the $v_{\rm flat}-M_{\rm halo}$ relation which implies that, at a given $v_{\rm flat}$, simulated discs inhabit more massive halos than real ones. 
The combination of these two effects produces the mismatch observed.

Simulated discs partially overlap with the data in the ($f_\star$,~$M_\star$) plane, but the comparison is limited as the former are very rare at $M_\star>10^{11}\msun$ (see Section \ref{discussion}).
Lowering the thresholds in $\mathcal{R}_\star$ and $\mathcal{F}_\star$ would allow for a larger sample of discs at higher $M_\star$ (grey squares in Fig.\,\ref{fig:shmr}), but these would not compare favourably with SPARC given that the observed and simulated populations diverge significantly at higher masses.
At $1\!<\!M_\star/\msun\!<\!3\times10^{11}$, the median $f_\star$ of EAGLE (IllustrisTNG) discs is $0.13$ ($0.23$), while in SPARC it is $0.48$, with some individual systems reaching unity.
These considerations highlight the difficulty of producing massive disc galaxies in numerical simulations with the observed global stellar-to-dark matter mass ratio.

\begin{figure*}
\begin{center}
\includegraphics[width=0.9\textwidth]{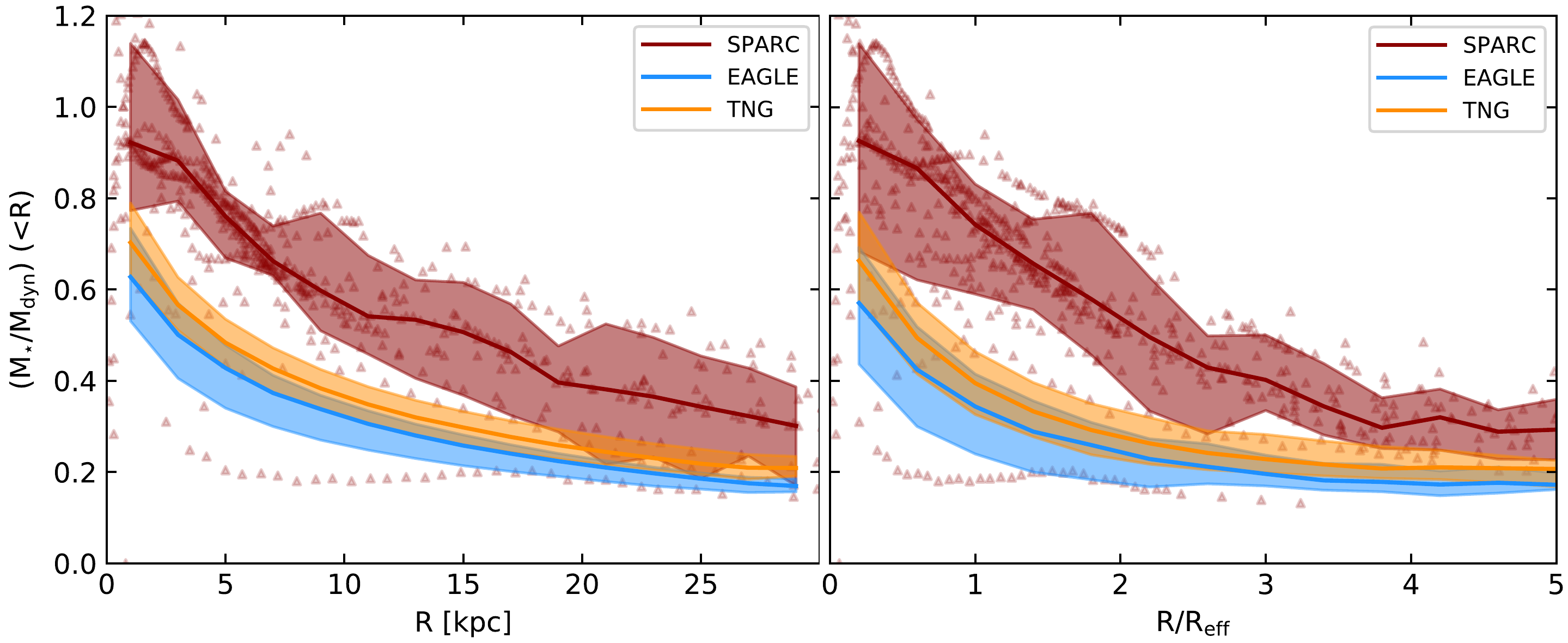}
\caption{\emph{Left panel}: Stellar-to-total enclosed mass profiles for massive disc galaxies in the EAGLE (blue) and IllustrisTNG (orange) simulations, compared with the data from SPARC (red). The solid lines show the median profiles, while the shaded areas represent the scatter given by the difference between the 84th and the 16th percentiles. Individual measurements for SPARC spirals are shown as red triangles. 
\emph{Right panel}: As in the left panel, but radii are normalised to the effective radius $R_{\rm eff}$ of each galaxy.}
\label{fig:Mstar_Mdyn}
\end{center}
\end{figure*}

\subsection{Local disc-halo connection} \label{ssec:SHMR_local}
Additional insights into the disc-halo connection can be obtained by studying the mass distribution at local scales, i.e. within the galaxy discs.
In Fig.\,\ref{fig:Mstar_Mdyn} we show how the ratio between the stellar mass $M_\star(R)$ and the total dynamical mass $M_{\rm dyn}(R)$ enclosed within a given radius $R$ varies as a function of $R$ for the massive discs in SPARC and in the simulations.
For SPARC galaxies we derive $M_\star(R)$ from their $3.6\um$ surface brightness profile assuming a razor-thin disc geometry and mass-to-light ratios from \citetalias{Posti+19}.
Provided that the contribution of the
gas to the mass budget within the disc is very small, we set $M_{\rm dyn}(R) \simeq M_\star(R) + R\,(v_{\rm obs}^2-v_\star^2)/G$.
In the simulations, instead, we compute enclosed masses directly form the particle data. 

 Fig.\,\ref{fig:Mstar_Mdyn} clearly shows that the simulations underestimate the contribution of stars to the total mass budget at all radii. 
The discrepancy is a factor of $\sim2$ between $1$ and $2\times$ the effective radius ($R_{\rm eff}$), and decreases down to a factor $1.5$ at smaller and larger radii.
Simulated discs already become dark matter dominated at $R\!\sim\!5\kpc$, while stars in SPARC spirals constitute the main dynamical component out to $R\sim15\kpc$.
This implies that not only simulated discs inhabit heavier halos than observed, but that also the internal dynamics of these discs on local scales is more dark matter-dominated than observed.
In this context, the offset in the size-$M_\star$ relation noticeable from Fig.\,\ref{fig:scaling_relations} plays an important role since, at a given $M_\star$, larger discs have lower stellar surface densities corresponding to lower radial acceleration at any given radius.
Furthermore, the concentrations inferred from a decomposition of the rotation curves of simulated galaxies (Appendix \ref{app:mass_dec}) are a factor of $\sim2$ higher than those determined for SPARC galaxies of the same inferred $M_{\rm halo}$. This drives down $M_\star(R)/M_{\rm dyn}(R)$ even further in the inner galactic regions of the simulated galaxies.

This local discrepancy is another important manifestation of the peculiar galaxy-halo connection of massive discs, which is coupled to the global discrepancy already noted by \citetalias{Posti+19}.
Our new findings on the local stellar-to-dynamical mass ratio are important in this context, since while one might argue that to solve the $f_\star$ discrepancy on global scales discs would need to inhabit even higher-concentration halos (so that the same circular velocity is obtained in less massive halos), this would further exacerbate the $(M_\star/M_{\rm dyn})(<R)$ discrepancy on local scales, ruling this out as a viable solution. Thus, all of our results combined suggest that, in order to explain their observed properties, massive spirals need to have \emph{everywhere} less dark matter than expected from AM models.

\section{Discussion and Summary}\label{discussion}
The relation between the stellar and the dark matter masses in nearby disc galaxies seem to be well described by a simple power law \citep{Posti+19b}, which translates into a monotonic relation between $M_\star$ and the star formation efficiency $f_\star$ \citepalias{Posti+19}.
As a consequence, massive ($M_\star\gtrsim5\times10^{10}\msun$) spirals depart significantly from the predictions of AM methods, reaching $f_\star$ of about unity at the high-$M_\star$ end (failed feedback problem).
This result is largely independent of the halo profile model adopted in the kinematic decomposition, as we did also verify using the various mass models provided by \citet{Ghari+19} and \citet{Li+20} for the SPARC dataset.

In this work we have analyzed this discrepancy, both on a global and local scale, comparing observations with predictions from two of the best-known recent cosmological hydrodynamical simulations, EAGLE and IllustrisTNG. 
These simulations feature an optimal compromise between box size and particle masses, which allows to sample at sub-kpc scale resolution several tens of discs in the interested range of stellar masses.
Also, the parameters of these models are explicitly tuned to reproduce several properties of the $z=0$ galaxy population, providing a solid starting point to carry out further investigations.
Our results show that simulated discs appear to inhabit overly massive dark matter halos, and that their dynamics, unlike real spirals, are everywhere dark matter dominated by a wide margin.

An important question is whether our results are unique to the EAGLE and IllustrisTNG runs studied here or can be generalised to other simulations that adopt different resolution and/or sub-grid schemes.
As already pointed out in other studies \citep{Scannapieco+12,Hopkins+18}, feedback implementation is likely to have a much larger impact on the properties of simulated galaxies than the resolution itself.
As an example, the EAGLE L025N0752 run has about twice the linear resolution of the L100N1504 run studied here, yet the stellar-to-halo mass relation is the same \citep[see Fig.\,8 in][]{Schaye+15}, simply because feedback has been re-calibrated to achieve this. 
On the other hand, the original Illustris suite \citep{Vogelsberger+14} is comparable with IllustrisTNG in terms of resolution, but the different feedback implementation leads to enormous differences in the galaxy mass profiles \citep[Figs.\,3 and \,4 in][]{Lovell+18}
Interestingly, the Magneticum simulations have a factor $\sim6$ lower mass resolution than the runs considered here and feature both discs and spheroids with quite large $f_{\star}$ \citep[up to $\sim0.5$,][]{Teklu+17}, suggesting that particle mass and $f_{\star}$ may be positively correlated.
Moving to zoom-in runs, in the NIHAO suite \citep{Wang+15} the mass resolution varies depending on the system mass - still remaining several times higher than EAGLE and IllustrisTNG for the most massive galaxies - yet the SHMR follows quite closely prediction from abundance matching, with maximum $f_\star$ of $\sim0.3$ \citep[Fig.\,5 in][]{Wang+15}. 
Similar values of $f_\star$ are found in the NIHAO-UHD runs \citep{Buck+20}.
The most massive ($M_\star\sim10^{11}\msun$) discs in the Latte/FIRE-2 simulations \citep{Hopkins+18} have a gas particle mass of $\sim10^4\msun$ and $f_\star$ between $0.3$ and $0.4$ \citep{Ma+18}, similar to what we find in IllustrisTNG. 
While we acknowledge the existence of a broad range of predictions in theoretical models, mostly originating from different implementations of the stellar and AGN feedback physics, we stress that the goal of the present work is not to offer a complete picture of the disc-halo connection in the vast realm of simulations, but rather to highlight the presence of important discrepancies between models and data using two complementary, well-studied suites which have been specifically designed to capture the properties of $z\!=\!0$ galaxies.

In the simulations we have computed stellar and halo masses using two different approaches: we have extracted them directly from the galaxy catalogues and have determined them via a mass-decomposition of circular velocity profiles.
The two methods lead to compatible results (see Fig.\,\ref{fig:true_vs_observed}), which incidentally validates the approach adopted by \citetalias{Posti+19} for their analysis.
However, rotation curves in SPARC are determined from \hi\ (and, in part, \ha) data, and one may argue that tracing the azimuthal speed of cold gas in the simulated galaxies, rather then extracting their $v_{\rm c}$ profile, would lead to a more direct comparison with the observations.
We have checked this using the approach described by \citet{Oman+19} to derive the \hi\ content of the gas particles in the simulations, and measured the \hi\ rotational speed in annuli of $1\kpc$ width oriented according to the \hi\ angular momentum vector of each galaxy.
Unfortunately, we found that the \hi\ kinematics of simulated discs is strongly disturbed (especially in EAGLE) and only in rare cases approaches the circular velocity.
The reasons for this remain to be clarified, but may well be related to disturbances induced by over-efficient feedback from star formation.

Another explanation for the lack of massive discs with high $f_\star$ in the models may be that the SPARC sample, which is not volume-limited, is made of rare, special systems that are not representative of the overall population of spirals.
A simple argument demonstrates that this is not the case: $13$ out of the $21$ high-$f_\star$ SPARC galaxies are located within a distance of $62\Mpc$, which encompasses a spherical volume equivalent to those of the runs studied here.
The systematic lack of these objects in the simulations must therefore be due to a deficiency of the models, rather than to a bias in the observations.

We have shown that the discrepancy in the stellar-to-dark matter ratio between simulated and observed systems is both global and local, and extends well into their inner regions where the former are dark-matter dominated at all radii (except for the innermost $\sim5\kpc$), while in the latter the stellar component dominates the galaxy dynamics.
The dynamical importance of stars in observed massive spirals is well known and emerges directly from the shape of their rotation curves, which follows very closely the light profile \citep[the so-called Renzo's rule,][]{Sancisi04} for several kpc before flattening out.
The small-scale discrepancy was already outlined by \citet{Ludlow+17} when investigating the radial acceleration relation\footnote{See also \citet{McGaugh+07} for a discussion of the mass discrepancy at all radii.} \citep{McGaugh+16} in EAGLE discs, by \citet{Lovell+18} for high-mass IllustrisTNG galaxies, and seems to be even stronger in other simulation suites such as SIMBA \citep{Dave+19,Glowacki+20}.
Here, we have highlighted how the problem exists at all scales.
Clearly, at a fixed $f_\star$, differences in the halo density profiles and in the disc sizes can lead to very diverse $M_\star/M_{\rm dyn}$ profiles, thus the relation between small and large scales is not trivially set by $f_\star$.
The simulated halos appear to be highly concentrated, probably as a result of halo contraction, which represents a plausible explanation for the discrepancy with observations in the regions of the discs.

It is possible to draw a parallel between these simulated discs and another category of dark matter-dominated objects: dwarf galaxies.
The dominance of dark matter in the simulated discs results in a poor variety in their circular velocity profiles (not shown in this work).
As the star formation efficiency of the host halos is nearly constant (Fig.\,\ref{fig:shmr}), this results in a very narrow TFR (Fig.\,\ref{fig:scaling_relations}) and little scatter in the $M_\star/M_{\rm dyn}$ profiles (Fig.\,\ref{fig:Mstar_Mdyn}).
In contrast, the observed galaxies seem to show more variety in their $f_\star$, rotation curves and $M_\star/M_{\rm dyn}$ profiles.
This echoes the `diversity problem' for dwarf galaxies \citep{Oman+15}, where the self-similar shape of $v_{\rm c}$ profiles in simulated dwarfs is in tension with the diversity in the \hi\ rotation curves that can be found in the observed population, at a fixed rotational speed.
There are however two important differences between the low- and high-mass regimes.
The first is that the \hi\ kinematics in dwarf galaxies is typically more disturbed \citep{Oh+15,Iorio+17}, and this results in more uncertain rotation curves which can be, to some extent, model-dependent \citep[e.g.][]{SpekkensSellwood07}.
The second is that the diversity problem in dwarfs is closely related to the cusp-core issue \citep{FloresPrimack94,Moore94}, which concerns the innermost central regions of dwarfs, whereas the discrepancy at the high-mass end is both local and global.

The results of \citetalias{Posti+19} suggest the existence of different pathways for the creation of massive late-type and early-type systems, with the former resulting from more gentle merging histories which would lead to lower dark matter content and higher star formation efficiencies \citep[although rejuvenation of ancient spheroids via gas-rich mergers is also a possibility, see][]{Jackson+20}.

We note that such a scenario is not inconsistent with the existence of a unique SHMR (with scatter), as predicted by current $\Lambda$CDM models.
As discussed by \citet{Moster+19}, active and passive systems are expected to distribute differently within the scatter of the SHMR due to the diverse accretion histories of their halos.
This effect, combined with the shallow slope of the SHMR at large masses, is such that at a fixed $M_\star$ active (passive) galaxies at $z\!=\!0$ are scattered preferentially towards lower (higher) halo masses, in line with the observed trend.
However, at a $M_\star$ of $10^{11}\msun$, this should lead to a difference of $\sim0.2$ dex in the $f_\star$ of active and passive galaxies, which is approximately what we find comparing our EAGLE and IllustrisTNG discs with the whole population of centrals at similar $M_\star$, but is largely insufficient to justify the difference between the AM prediction and the observed data.

It is possible, though, that a larger scatter in the SHMR may help in reconciling observations and theory.
Indeed, the recent discovery of super-spirals \citep{Ogle+19a}, which are extremely massive ($\log(M_\star/\msun)>11.5$) late-type systems with rotational speed up to $\sim570\kms$ \citep{Ogle+19b}, suggests that the population of high-$M_\star$ disc galaxies is more heterogeneous than previously thought.
Such systems must be rare, as only $3$ objects with \hi\ velocity width (W$_{50}$) larger than $800\kms$ are present in the ALFALFA-100 catalogue \citep{Haynes+18}.
Yet, their existence is symptomatic of a wider distribution of feedback efficiencies at fixed halo mass.
In the specific case of the massive spirals considered here, less efficient feedback from stars and/or AGNs can help in bringing the models closer to the data on both global and local scales, leading to higher stellar masses at fixed halo mass and to smaller disc sizes due to less efficient angular momentum redistribution from the outer to the inner regions of the halo's gas reservoir \citep[e.g.][]{Brook+12}.

While a lower feedback efficiency may alleviate the problems discussed in this work, the way it would affect other galaxy properties remains to be clarified.
For instance, different theoretical models of galaxy evolution give very different predictions for the growth of discs in the absence of galaxy-scale, powerful feedback episodes \citep[e.g.][]{NavarroSteinmetz97,Pezzulli+17}.
More importantly, the increase of $f_\star$ is very likely to have an impact on the dark matter distribution by promoting halo contraction \citep{Blumenthal+84,Gnedin+04,Dutton+16}, which may exacerbate the local problem discussed here by lowering the $M_\star/M_{\rm dyn}$ profile in the inner regions \citep[although halo expansion associated to higher $f_\star$ is sometimes seen in simulations, see][]{Lovell+18}.
This local problem might be considered as a residual, restricted to massive discs, of the old problem pointed out by \citet{NavarroSteinmetz00} who already found overly-concentrated dark matter halos in early hydrodynamical simulations of Milky Way-like systems.
In this context, the solution for the local and global discrepancies presented in this work hinges on whether a feedback recipe that increases the star formation efficiency in $\sim10^{12}\msun$ halos while simultaneously avoiding central over-condensation and excessive halo contraction is achievable.
Future generations of hydrodynamical simulations of galaxy formation may help to clarify this point.

In summary, our findings indicate that the population of high-mass spirals emerging from state-of-the-art $\Lambda$CDM cosmological simulations EAGLE and IllustrisTNG differs systematically from that which we observe in terms of both local and global stellar-to-dark matter content.
The difference cannot be explained in terms of selection effects or limitations in the modelling of the observed data, and clearly points to a mismatch in the efficiency of massive spirals at converting their baryons into stars.
Ultimately, the difference can be understood in terms of a more scattered galaxy-halo connection at the high end of the stellar mass function, which, at the current stage, does not seem to emerge from numerical models in $\Lambda$CDM framework.

\begin{acknowledgements}
The authors thank the referee, Florent Renaud, for his careful and thoughtful report.
AM and GC acknowledge the support by INAF/Frontiera through the "Progetti Premiali" funding scheme of the Italian Ministry of Education, University, and Research.
BF acknowledges funding from the Agence Nationale de la Recherche (ANR project ANR-18-CE31-0006 and ANR-19-CE31-0017) and from the European Research Council (ERC) under the European Union's Horizon 2020 research and innovation programme (grant agreement No. 834148).
LP acknowledges support from the Centre National d’Etudes Spatiales (CNES). KO acknowledges support by the European Research Council (ERC) through Advanced Investigator grant to C.S. Frenk, DMIDAS (GA 786910).
\end{acknowledgements}


\bibliographystyle{aa} 
\bibliography{shmr_sims} 

\begin{thebibliography}{70}
\expandafter\ifx\csname natexlab\endcsname\relax\def\natexlab#1{#1}\fi

\bibitem[{{Bah{\'e}} {et~al.}(2016){Bah{\'e}}, {Crain}, {Kauffmann}, {Bower},
  {Schaye}, {Furlong}, {Lagos}, {Schaller}, {Trayford}, {Dalla Vecchia}, \&
  {Theuns}}]{Bahe+16}
{Bah{\'e}}, Y.~M., {Crain}, R.~A., {Kauffmann}, G., {et~al.} 2016, \mnras, 456,
  1115

\bibitem[{{Behroozi} {et~al.}(2010){Behroozi}, {Conroy}, \&
  {Wechsler}}]{Behroozi+10}
{Behroozi}, P.~S., {Conroy}, C., \& {Wechsler}, R.~H. 2010, \apj, 717, 379

\bibitem[{{Blumenthal} {et~al.}(1984){Blumenthal}, {Faber}, {Primack}, \&
  {Rees}}]{Blumenthal+84}
{Blumenthal}, G.~R., {Faber}, S.~M., {Primack}, J.~R., \& {Rees}, M.~J. 1984,
  \nat, 311, 517

\bibitem[{{Brook} {et~al.}(2012){Brook}, {Stinson}, {Gibson}, {Ro{\v{s}}kar},
  {Wadsley}, \& {Quinn}}]{Brook+12}
{Brook}, C.~B., {Stinson}, G., {Gibson}, B.~K., {et~al.} 2012, \mnras, 419, 771

\bibitem[{{Buck} {et~al.}(2020){Buck}, {Obreja}, {Macci{\`o}}, {Minchev},
  {Dutton}, \& {Ostriker}}]{Buck+20}
{Buck}, T., {Obreja}, A., {Macci{\`o}}, A.~V., {et~al.} 2020, \mnras, 491, 3461

\bibitem[{{Cappellari} {et~al.}(2013){Cappellari}, {Scott}, {Alatalo}, {Blitz},
  {Bois}, {Bournaud}, {Bureau}, {Crocker}, {Davies}, {Davis}, {de Zeeuw},
  {Duc}, {Emsellem}, {Khochfar}, {Krajnovi{\'c}}, {Kuntschner}, {McDermid},
  {Morganti}, {Naab}, {Oosterloo}, {Sarzi}, {Serra}, {Weijmans}, \&
  {Young}}]{Cappellari+13}
{Cappellari}, M., {Scott}, N., {Alatalo}, K., {et~al.} 2013, \mnras, 432, 1709

\bibitem[{{Crain} {et~al.}(2015){Crain}, {Schaye}, {Bower}, {Furlong},
  {Schaller}, {Theuns}, {Dalla Vecchia}, {Frenk}, {McCarthy}, {Helly},
  {Jenkins}, {Rosas-Guevara}, {White}, \& {Trayford}}]{Crain+15}
{Crain}, R.~A., {Schaye}, J., {Bower}, R.~G., {et~al.} 2015, \mnras, 450, 1937

\bibitem[{{Dav{\'e}} {et~al.}(2019){Dav{\'e}}, {Angl{\'e}s-Alc{\'a}zar},
  {Narayanan}, {Li}, {Rafieferantsoa}, \& {Appleby}}]{Dave+19}
{Dav{\'e}}, R., {Angl{\'e}s-Alc{\'a}zar}, D., {Narayanan}, D., {et~al.} 2019,
  \mnras, 486, 2827

\bibitem[{{Diemer} {et~al.}(2019){Diemer}, {Stevens}, {Lagos}, {Calette},
  {Tacchella}, {Hernquist}, {Marinacci}, {Nelson}, {Pillepich},
  {Rodriguez-Gomez}, {Villaescusa-Navarro}, \& {Vogelsberger}}]{Diemer+19}
{Diemer}, B., {Stevens}, A. R.~H., {Lagos}, C. d.~P., {et~al.} 2019, \mnras,
  487, 1529

\bibitem[{{Driver} {et~al.}(2011){Driver}, {Hill}, {Kelvin}, {Robotham},
  {Liske}, {Norberg}, {Baldry}, {Bamford}, {Hopkins}, {Loveday}, {Peacock},
  {Andrae}, {Bland -Hawthorn}, {Brough}, {Brown}, {Cameron}, {Ching},
  {Colless}, {Conselice}, {Croom}, {Cross}, {de Propris}, {Dye}, {Drinkwater},
  {Ellis}, {Graham}, {Grootes}, {Gunawardhana}, {Jones}, {van Kampen},
  {Maraston}, {Nichol}, {Parkinson}, {Phillipps}, {Pimbblet}, {Popescu},
  {Prescott}, {Roseboom}, {Sadler}, {Sansom}, {Sharp}, {Smith}, {Taylor},
  {Thomas}, {Tuffs}, {Wijesinghe}, {Dunne}, {Frenk}, {Jarvis}, {Madore},
  {Meyer}, {Seibert}, {Staveley-Smith}, {Sutherland}, \& {Warren}}]{Driver+11}
{Driver}, S.~P., {Hill}, D.~T., {Kelvin}, L.~S., {et~al.} 2011, \mnras, 413,
  971

\bibitem[{{Dutton} {et~al.}(2010){Dutton}, {Conroy}, {van den Bosch}, {Prada},
  \& {More}}]{Dutton+10}
{Dutton}, A.~A., {Conroy}, C., {van den Bosch}, F.~C., {Prada}, F., \& {More},
  S. 2010, \mnras, 407, 2

\bibitem[{{Dutton} \& {Macci{\`o}}(2014)}]{DuttonMaccio14}
{Dutton}, A.~A. \& {Macci{\`o}}, A.~V. 2014, \mnras, 441, 3359

\bibitem[{{Dutton} {et~al.}(2016){Dutton}, {Macci{\`o}}, {Dekel}, {Wang},
  {Stinson}, {Obreja}, {Di Cintio}, {Brook}, {Buck}, \& {Kang}}]{Dutton+16}
{Dutton}, A.~A., {Macci{\`o}}, A.~V., {Dekel}, A., {et~al.} 2016, \mnras, 461,
  2658

\bibitem[{{Flores} \& {Primack}(1994)}]{FloresPrimack94}
{Flores}, R.~A. \& {Primack}, J.~R. 1994, \apjl, 427, L1

\bibitem[{{Ghari} {et~al.}(2019){Ghari}, {Famaey}, {Laporte}, \&
  {Haghi}}]{Ghari+19}
{Ghari}, A., {Famaey}, B., {Laporte}, C., \& {Haghi}, H. 2019, \aap, 623, A123

\bibitem[{Glowacki {et~al.}(2020)Glowacki, Elson, \& Davé}]{Glowacki+20}
Glowacki, M., Elson, E., \& Davé, R. 2020, The baryonic Tully-Fisher relation
  in the Simba simulation

\bibitem[{{Gnedin} {et~al.}(2004){Gnedin}, {Kravtsov}, {Klypin}, \&
  {Nagai}}]{Gnedin+04}
{Gnedin}, O.~Y., {Kravtsov}, A.~V., {Klypin}, A.~A., \& {Nagai}, D. 2004, \apj,
  616, 16

\bibitem[{{Haynes} {et~al.}(2018){Haynes}, {Giovanelli}, {Kent}, {Adams},
  {Balonek}, {Craig}, {Fertig}, {Finn}, {Giovanardi}, {Hallenbeck}, {Hess},
  {Hoffman}, {Huang}, {Jones}, {Koopmann}, {Kornreich}, {Leisman}, {Miller},
  {Moorman}, {O'Connor}, {O'Donoghue}, {Papastergis}, {Troischt}, {Stark}, \&
  {Xiao}}]{Haynes+18}
{Haynes}, M.~P., {Giovanelli}, R., {Kent}, B.~R., {et~al.} 2018, \apj, 861, 49

\bibitem[{{Hopkins} {et~al.}(2018){Hopkins}, {Wetzel}, {Kere{\v{s}}},
  {Faucher-Gigu{\`e}re}, {Quataert}, {Boylan-Kolchin}, {Murray}, {Hayward},
  {Garrison-Kimmel}, {Hummels}, {Feldmann}, {Torrey}, {Ma},
  {Angl{\'e}s-Alc{\'a}zar}, {Su}, {Orr}, {Schmitz}, {Escala}, {Sanderson},
  {Grudi{\'c}}, {Hafen}, {Kim}, {Fitts}, {Bullock}, {Wheeler}, {Chan},
  {Elbert}, \& {Narayanan}}]{Hopkins+18}
{Hopkins}, P.~F., {Wetzel}, A., {Kere{\v{s}}}, D., {et~al.} 2018, \mnras, 480,
  800

\bibitem[{{Iorio} {et~al.}(2017){Iorio}, {Fraternali}, {Nipoti}, {Di Teodoro},
  {Read}, \& {Battaglia}}]{Iorio+17}
{Iorio}, G., {Fraternali}, F., {Nipoti}, C., {et~al.} 2017, \mnras, 466, 4159

\bibitem[{{Jackson} {et~al.}(2020){Jackson}, {Martin}, {Kaviraj}, {Laigle},
  {Devriendt}, {Dubois}, \& {Pichon}}]{Jackson+20}
{Jackson}, R.~A., {Martin}, G., {Kaviraj}, S., {et~al.} 2020, \mnras
  [\eprint[arXiv]{2004.00023}]

\bibitem[{{Kravtsov} {et~al.}(2018){Kravtsov}, {Vikhlinin}, \&
  {Meshcheryakov}}]{Kravtsov+18}
{Kravtsov}, A.~V., {Vikhlinin}, A.~A., \& {Meshcheryakov}, A.~V. 2018,
  Astronomy Letters, 44, 8

\bibitem[{{Leauthaud} {et~al.}(2012){Leauthaud}, {Tinker}, {Bundy}, {Behroozi},
  {Massey}, {Rhodes}, {George}, {Kneib}, {Benson}, {Wechsler}, {Busha},
  {Capak}, {Cort{\^e}s}, {Ilbert}, {Koekemoer}, {Le F{\`e}vre}, {Lilly},
  {McCracken}, {Salvato}, {Schrabback}, {Scoville}, {Smith}, \&
  {Taylor}}]{Leauthaud+12}
{Leauthaud}, A., {Tinker}, J., {Bundy}, K., {et~al.} 2012, \apj, 744, 159

\bibitem[{{Lelli} {et~al.}(2016){Lelli}, {McGaugh}, \& {Schombert}}]{Lelli+16}
{Lelli}, F., {McGaugh}, S.~S., \& {Schombert}, J.~M. 2016, \aj, 152, 157

\bibitem[{{Li} {et~al.}(2020){Li}, {Lelli}, {McGaugh}, \& {Schombert}}]{Li+20}
{Li}, P., {Lelli}, F., {McGaugh}, S., \& {Schombert}, J. 2020, \apjs, 247, 31

\bibitem[{{Lovell} {et~al.}(2018){Lovell}, {Pillepich}, {Genel}, {Nelson},
  {Springel}, {Pakmor}, {Marinacci}, {Weinberger}, {Torrey}, {Vogelsberger},
  {Alabi}, \& {Hernquist}}]{Lovell+18}
{Lovell}, M.~R., {Pillepich}, A., {Genel}, S., {et~al.} 2018, \mnras, 481, 1950

\bibitem[{{Ludlow} {et~al.}(2017){Ludlow}, {Ben{\'\i}tez-Llambay}, {Schaller},
  {Theuns}, {Frenk}, {Bower}, {Schaye}, {Crain}, {Navarro}, {Fattahi}, \&
  {Oman}}]{Ludlow+17}
{Ludlow}, A.~D., {Ben{\'\i}tez-Llambay}, A., {Schaller}, M., {et~al.} 2017,
  \prl, 118, 161103

\bibitem[{{Ma} {et~al.}(2018){Ma}, {Hopkins}, {Garrison-Kimmel},
  {Faucher-Gigu{\`e}re}, {Quataert}, {Boylan-Kolchin}, {Hayward}, {Feldmann},
  \& {Kere{\v{s}}}}]{Ma+18}
{Ma}, X., {Hopkins}, P.~F., {Garrison-Kimmel}, S., {et~al.} 2018, \mnras, 478,
  1694

\bibitem[{{Mandelbaum} {et~al.}(2006){Mandelbaum}, {Seljak}, {Kauffmann},
  {Hirata}, \& {Brinkmann}}]{Mandelbaum+06}
{Mandelbaum}, R., {Seljak}, U., {Kauffmann}, G., {Hirata}, C.~M., \&
  {Brinkmann}, J. 2006, \mnras, 368, 715

\bibitem[{{Mandelbaum} {et~al.}(2016){Mandelbaum}, {Wang}, {Zu}, {White},
  {Henriques}, \& {More}}]{Mandelbaum+16}
{Mandelbaum}, R., {Wang}, W., {Zu}, Y., {et~al.} 2016, \mnras, 457, 3200

\bibitem[{{McAlpine} {et~al.}(2016){McAlpine}, {Helly}, {Schaller}, {Trayford},
  {Qu}, {Furlong}, {Bower}, {Crain}, {Schaye}, {Theuns}, {Dalla Vecchia},
  {Frenk}, {McCarthy}, {Jenkins}, {Rosas-Guevara}, {White}, {Baes}, {Camps}, \&
  {Lemson}}]{McAlpine+16}
{McAlpine}, S., {Helly}, J.~C., {Schaller}, M., {et~al.} 2016, Astronomy and
  Computing, 15, 72

\bibitem[{{McGaugh} {et~al.}(2007){McGaugh}, {de Blok}, {Schombert}, {Kuzio de
  Naray}, \& {Kim}}]{McGaugh+07}
{McGaugh}, S.~S., {de Blok}, W.~J.~G., {Schombert}, J.~M., {Kuzio de Naray},
  R., \& {Kim}, J.~H. 2007, \apj, 659, 149

\bibitem[{{McGaugh} {et~al.}(2016){McGaugh}, {Lelli}, \&
  {Schombert}}]{McGaugh+16}
{McGaugh}, S.~S., {Lelli}, F., \& {Schombert}, J.~M. 2016, \prl, 117, 201101

\bibitem[{{Moffett} {et~al.}(2016){Moffett}, {Ingarfield}, {Driver},
  {Robotham}, {Kelvin}, {Lange}, {Me{\v{s}}tri{\'c}}, {Alpaslan}, {Baldry},
  {Bland-Hawthorn}, {Brough}, {Cluver}, {Davies}, {Holwerda}, {Hopkins},
  {Kafle}, {Kennedy}, {Norberg}, \& {Taylor}}]{Moffett+16}
{Moffett}, A.~J., {Ingarfield}, S.~A., {Driver}, S.~P., {et~al.} 2016, \mnras,
  457, 1308

\bibitem[{{Moore}(1994)}]{Moore94}
{Moore}, B. 1994, \nat, 370, 629

\bibitem[{{More} {et~al.}(2011){More}, {van den Bosch}, {Cacciato}, {Skibba},
  {Mo}, \& {Yang}}]{More+11}
{More}, S., {van den Bosch}, F.~C., {Cacciato}, M., {et~al.} 2011, \mnras, 410,
  210

\bibitem[{{Moster} {et~al.}(2013){Moster}, {Naab}, \& {White}}]{Moster+13}
{Moster}, B.~P., {Naab}, T., \& {White}, S. D.~M. 2013, \mnras, 428, 3121

\bibitem[{{Moster} {et~al.}(2019){Moster}, {Naab}, \& {White}}]{Moster+19}
{Moster}, B.~P., {Naab}, T., \& {White}, S. D.~M. 2019, arXiv e-prints,
  arXiv:1910.09552

\bibitem[{{Navarro} {et~al.}(1996){Navarro}, {Frenk}, \& {White}}]{NFW}
{Navarro}, J.~F., {Frenk}, C.~S., \& {White}, S. D.~M. 1996, \apj, 462, 563

\bibitem[{{Navarro} \& {Steinmetz}(1997)}]{NavarroSteinmetz97}
{Navarro}, J.~F. \& {Steinmetz}, M. 1997, \apj, 478, 13

\bibitem[{{Navarro} \& {Steinmetz}(2000)}]{NavarroSteinmetz00}
{Navarro}, J.~F. \& {Steinmetz}, M. 2000, \apj, 528, 607

\bibitem[{{Nelson} {et~al.}(2019){Nelson}, {Springel}, {Pillepich},
  {Rodriguez-Gomez}, {Torrey}, {Genel}, {Vogelsberger}, {Pakmor}, {Marinacci},
  {Weinberger}, {Kelley}, {Lovell}, {Diemer}, \& {Hernquist}}]{Nelson+19}
{Nelson}, D., {Springel}, V., {Pillepich}, A., {et~al.} 2019, Computational
  Astrophysics and Cosmology, 6, 2

\bibitem[{{Ogle} {et~al.}(2019{\natexlab{a}}){Ogle}, {Jarrett}, {Lanz},
  {Cluver}, {Alatalo}, {Appleton}, \& {Mazzarella}}]{Ogle+19b}
{Ogle}, P.~M., {Jarrett}, T., {Lanz}, L., {et~al.} 2019{\natexlab{a}}, \apjl,
  884, L11

\bibitem[{{Ogle} {et~al.}(2019{\natexlab{b}}){Ogle}, {Lanz}, {Appleton},
  {Helou}, \& {Mazzarella}}]{Ogle+19a}
{Ogle}, P.~M., {Lanz}, L., {Appleton}, P.~N., {Helou}, G., \& {Mazzarella}, J.
  2019{\natexlab{b}}, \apjs, 243, 14

\bibitem[{{Oh} {et~al.}(2015){Oh}, {Hunter}, {Brinks}, {Elmegreen}, {Schruba},
  {Walter}, {Rupen}, {Young}, {Simpson}, {Johnson}, {Herrmann}, {Ficut-Vicas},
  {Cigan}, {Heesen}, {Ashley}, \& {Zhang}}]{Oh+15}
{Oh}, S.-H., {Hunter}, D.~A., {Brinks}, E., {et~al.} 2015, \aj, 149, 180

\bibitem[{{Oman} {et~al.}(2019){Oman}, {Marasco}, {Navarro}, {Frenk}, {Schaye},
  \& {Ben{\'\i}tez-Llambay}}]{Oman+19}
{Oman}, K.~A., {Marasco}, A., {Navarro}, J.~F., {et~al.} 2019, \mnras, 482, 821

\bibitem[{{Oman} {et~al.}(2015){Oman}, {Navarro}, {Fattahi}, {Frenk}, {Sawala},
  {White}, {Bower}, {Crain}, {Furlong}, {Schaller}, {Schaye}, \&
  {Theuns}}]{Oman+15}
{Oman}, K.~A., {Navarro}, J.~F., {Fattahi}, A., {et~al.} 2015, \mnras, 452,
  3650

\bibitem[{{Persic} {et~al.}(1996){Persic}, {Salucci}, \& {Stel}}]{Persic+96}
{Persic}, M., {Salucci}, P., \& {Stel}, F. 1996, \mnras, 281, 27

\bibitem[{{Pezzulli} {et~al.}(2017){Pezzulli}, {Fraternali}, \&
  {Binney}}]{Pezzulli+17}
{Pezzulli}, G., {Fraternali}, F., \& {Binney}, J. 2017, \mnras, 467, 311

\bibitem[{{Pillepich} {et~al.}(2018){Pillepich}, {Springel}, {Nelson}, {Genel},
  {Naiman}, {Pakmor}, {Hernquist}, {Torrey}, {Vogelsberger}, {Weinberger}, \&
  {Marinacci}}]{Pillepich+18}
{Pillepich}, A., {Springel}, V., {Nelson}, D., {et~al.} 2018, \mnras, 473, 4077

\bibitem[{{Planck Collaboration} {et~al.}(2018){Planck Collaboration},
  {Aghanim}, {Akrami}, {Ashdown}, {Aumont}, {Baccigalupi}, {Ballardini},
  {Banday}, {Barreiro}, {Bartolo}, {Basak}, {Battye}, {Benabed}, {Bernard},
  {Bersanelli}, {Bielewicz}, {Bock}, {Bond}, {Borrill}, {Bouchet}, {Boulanger},
  {Bucher}, {Burigana}, {Butler}, {Calabrese}, {Cardoso}, {Carron},
  {Challinor}, {Chiang}, {Chluba}, {Colombo}, {Combet}, {Contreras}, {Crill},
  {Cuttaia}, {de Bernardis}, {de Zotti}, {Delabrouille}, {Delouis}, {Di
  Valentino}, {Diego}, {Dor{\'e}}, {Douspis}, {Ducout}, {Dupac}, {Dusini},
  {Efstathiou}, {Elsner}, {En{\ss}lin}, {Eriksen}, {Fantaye}, {Farhang},
  {Fergusson}, {Fernandez-Cobos}, {Finelli}, {Forastieri}, {Frailis},
  {Fraisse}, {Franceschi}, {Frolov}, {Galeotta}, {Galli}, {Ganga},
  {G{\'e}nova-Santos}, {Gerbino}, {Ghosh}, {Gonz{\'a}lez-Nuevo}, {G{\'o}rski},
  {Gratton}, {Gruppuso}, {Gudmundsson}, {Hamann}, {Handley}, {Hansen},
  {Herranz}, {Hildebrandt}, {Hivon}, {Huang}, {Jaffe}, {Jones}, {Karakci},
  {Keih{\"a}nen}, {Keskitalo}, {Kiiveri}, {Kim}, {Kisner}, {Knox},
  {Krachmalnicoff}, {Kunz}, {Kurki-Suonio}, {Lagache}, {Lamarre}, {Lasenby},
  {Lattanzi}, {Lawrence}, {Le Jeune}, {Lemos}, {Lesgourgues}, {Levrier},
  {Lewis}, {Liguori}, {Lilje}, {Lilley}, {Lindholm}, {L{\'o}pez-Caniego},
  {Lubin}, {Ma}, {Mac{\'\i}as-P{\'e}rez}, {Maggio}, {Maino}, {Mandolesi},
  {Mangilli}, {Marcos-Caballero}, {Maris}, {Martin}, {Martinelli},
  {Mart{\'\i}nez-Gonz{\'a}lez}, {Matarrese}, {Mauri}, {McEwen}, {Meinhold},
  {Melchiorri}, {Mennella}, {Migliaccio}, {Millea}, {Mitra},
  {Miville-Desch{\^e}nes}, {Molinari}, {Montier}, {Morgante}, {Moss}, {Natoli},
  {N{\o}rgaard-Nielsen}, {Pagano}, {Paoletti}, {Partridge}, {Patanchon},
  {Peiris}, {Perrotta}, {Pettorino}, {Piacentini}, {Polastri}, {Polenta},
  {Puget}, {Rachen}, {Reinecke}, {Remazeilles}, {Renzi}, {Rocha}, {Rosset},
  {Roudier}, {Rubi{\~n}o-Mart{\'\i}n}, {Ruiz-Granados}, {Salvati}, {Sandri},
  {Savelainen}, {Scott}, {Shellard}, {Sirignano}, {Sirri}, {Spencer},
  {Sunyaev}, {Suur-Uski}, {Tauber}, {Tavagnacco}, {Tenti}, {Toffolatti},
  {Tomasi}, {Trombetti}, {Valenziano}, {Valiviita}, {Van Tent}, {Vibert},
  {Vielva}, {Villa}, {Vittorio}, {Wand elt}, {Wehus}, {White}, {White},
  {Zacchei}, \& {Zonca}}]{Planck+18}
{Planck Collaboration}, {Aghanim}, N., {Akrami}, Y., {et~al.} 2018, arXiv
  e-prints, arXiv:1807.06209

\bibitem[{{Posti} {et~al.}(2019{\natexlab{a}}){Posti}, {Fraternali}, \&
  {Marasco}}]{Posti+19}
{Posti}, L., {Fraternali}, F., \& {Marasco}, A. 2019{\natexlab{a}}, \aap, 626,
  A56

\bibitem[{{Posti} {et~al.}(2019{\natexlab{b}}){Posti}, {Marasco}, {Fraternali},
  \& {Famaey}}]{Posti+19b}
{Posti}, L., {Marasco}, A., {Fraternali}, F., \& {Famaey}, B.
  2019{\natexlab{b}}, \aap, 629, A59

\bibitem[{{Read} {et~al.}(2017){Read}, {Iorio}, {Agertz}, \&
  {Fraternali}}]{Read+17}
{Read}, J.~I., {Iorio}, G., {Agertz}, O., \& {Fraternali}, F. 2017, \mnras,
  467, 2019

\bibitem[{{Sancisi}(2004)}]{Sancisi04}
{Sancisi}, R. 2004, in IAU Symposium, Vol. 220, Dark Matter in Galaxies, ed.
  S.~{Ryder}, D.~{Pisano}, M.~{Walker}, \& K.~{Freeman}, 233

\bibitem[{{Scannapieco} {et~al.}(2012){Scannapieco}, {Wadepuhl}, {Parry},
  {Navarro}, {Jenkins}, {Springel}, {Teyssier}, {Carlson}, {Couchman}, {Crain},
  {Dalla Vecchia}, {Frenk}, {Kobayashi}, {Monaco}, {Murante}, {Okamoto},
  {Quinn}, {Schaye}, {Stinson}, {Theuns}, {Wadsley}, {White}, \&
  {Woods}}]{Scannapieco+12}
{Scannapieco}, C., {Wadepuhl}, M., {Parry}, O.~H., {et~al.} 2012, \mnras, 423,
  1726

\bibitem[{{Schaye} {et~al.}(2015){Schaye}, {Crain}, {Bower}, {Furlong},
  {Schaller}, {Theuns}, {Dalla Vecchia}, {Frenk}, {McCarthy}, {Helly},
  {Jenkins}, {Rosas-Guevara}, {White}, {Baes}, {Booth}, {Camps}, {Navarro},
  {Qu}, {Rahmati}, {Sawala}, {Thomas}, \& {Trayford}}]{Schaye+15}
{Schaye}, J., {Crain}, R.~A., {Bower}, R.~G., {et~al.} 2015, \mnras, 446, 521

\bibitem[{{Spekkens} \& {Sellwood}(2007)}]{SpekkensSellwood07}
{Spekkens}, K. \& {Sellwood}, J.~A. 2007, \apj, 664, 204

\bibitem[{{Springel}(2005)}]{gadget2}
{Springel}, V. 2005, \mnras, 364, 1105

\bibitem[{{Springel}(2010)}]{Springel10}
{Springel}, V. 2010, \mnras, 401, 791

\bibitem[{{Teklu} {et~al.}(2017){Teklu}, {Remus}, {Dolag}, \&
  {Burkert}}]{Teklu+17}
{Teklu}, A.~F., {Remus}, R.-S., {Dolag}, K., \& {Burkert}, A. 2017, \mnras,
  472, 4769

\bibitem[{{Thob} {et~al.}(2019){Thob}, {Crain}, {McCarthy}, {Schaller},
  {Lagos}, {Schaye}, {Talens}, {James}, {Theuns}, \& {Bower}}]{Thob+19}
{Thob}, A. C.~R., {Crain}, R.~A., {McCarthy}, I.~G., {et~al.} 2019, \mnras,
  485, 972

\bibitem[{{Tully} \& {Fisher}(1977)}]{TullyFisher77}
{Tully}, R.~B. \& {Fisher}, J.~R. 1977, \aap, 500, 105

\bibitem[{{Vale} \& {Ostriker}(2004)}]{ValeOstriker04}
{Vale}, A. \& {Ostriker}, J.~P. 2004, \mnras, 353, 189

\bibitem[{{van den Bosch} {et~al.}(2019){van den Bosch}, {Lange}, \&
  {Zentner}}]{Bosch+19}
{van den Bosch}, F.~C., {Lange}, J.~U., \& {Zentner}, A.~R. 2019, \mnras, 488,
  4984

\bibitem[{{van den Bosch} {et~al.}(2004){van den Bosch}, {Norberg}, {Mo}, \&
  {Yang}}]{Bosch+04}
{van den Bosch}, F.~C., {Norberg}, P., {Mo}, H.~J., \& {Yang}, X. 2004, \mnras,
  352, 1302

\bibitem[{{Vogelsberger} {et~al.}(2014){Vogelsberger}, {Genel}, {Springel},
  {Torrey}, {Sijacki}, {Xu}, {Snyder}, {Nelson}, \&
  {Hernquist}}]{Vogelsberger+14}
{Vogelsberger}, M., {Genel}, S., {Springel}, V., {et~al.} 2014, \mnras, 444,
  1518

\bibitem[{{Wang} {et~al.}(2015){Wang}, {Dutton}, {Stinson}, {Macci{\`o}},
  {Penzo}, {Kang}, {Keller}, \& {Wadsley}}]{Wang+15}
{Wang}, L., {Dutton}, A.~A., {Stinson}, G.~S., {et~al.} 2015, \mnras, 454, 83

\bibitem[{{Wechsler} \& {Tinker}(2018)}]{WechslerTinker18}
{Wechsler}, R.~H. \& {Tinker}, J.~L. 2018, \araa, 56, 435

\bibitem[{{White} \& {Rees}(1978)}]{WhiteRees78}
{White}, S.~D.~M. \& {Rees}, M.~J. 1978, \mnras, 183, 341

\end{thebibliography}

\appendix
\section{Supplementary material}\label{app:supplementary}
Tables \ref{tab:catalogue_sim} and \ref{tab:catalogue_obs} list the main properties of the samples of simulated and observed massive disc galaxies studied in this work.
These tables are available their entirety in machine-readable form at \url{https://drive.google.com/file/d/1qksA7yKmzQHRJQ3U3gXexWOe0hW9RnVs/view?usp=sharing}.
\begin{table*}
\caption{Main properties of the sample of simulated massive discs studied in this work.}
\label{tab:catalogue_sim} 
\centering
\begin{tabular}{lccccccc}
\hline\hline\noalign{\vspace{5pt}}
 Simulation     & Galaxy ID    & $\log_{10}(M_\star/M_\odot)$ & $\log_{10}(M_{\rm halo}/M_\odot)$  & $v_{\rm flat}$ & $R_{\rm eff}$ & $\mathcal{R}_\star$  & $\mathcal{F}_\star$ \\
      (1)       &   (2)     &            (3)               &         (4)                        &        (5)     &        (6)    &     (7)              & (8)                \\
\noalign{\smallskip}
\hline\noalign{\vspace{5pt}}
EAGLE   &   14582105    &    $10.95$    &    $12.93$    &   $236.66$   &      $9.17$    &     $2.41$   &      $0.84$\\
EAGLE  &   14202038    &    $10.98$     &   $12.70$ &      $253.67$   &      $9.24$ &        $2.50$   &      $0.83$\\
EAGLE   &   15518507    &    $11.01$     &   $12.74$ &      $287.68$   &     $13.45$ &        $2.32$   &      $0.79$\\
    &       &         &   \dots &         &      &           &      \\
\hline\noalign{\vspace{3pt}}
IllustrisTNG  &      351452     &   $11.11$    &    $12.48$   &    $266.92$    &     $9.20$   &      $1.75$   &      $0.82$\\
IllustrisTNG  &      368436     &   $11.33$    &    $12.64$   &    $290.15$    &    $16.58$   &      $1.79$   &      $0.84$\\
IllustrisTNG  &      369366     &   $10.89$    &    $12.25$   &    $225.96$    &    $13.18$   &      $1.82$   &      $0.89$\\
    &       &         &   \dots &         &      &           &      \\
\noalign{\vspace{2pt}}\hline
\noalign{\vspace{5pt}}

\multicolumn{8}{p{0.8\textwidth}}{\textbf{Notes.} 
(1) Simulation suite, the runs analysed are Ref-L0100N1504 in EAGLE and TNG100-1 in IllustrisTNG; (2) galaxy ID from the catalogues of \citet{McAlpine+16} and \citet{Nelson+19}; (3)-(4) stellar and halo masses; (5) Velocity of the flat part of the rotation curve in $\kms$, defined as specified in Section \ref{method}; (6) effective (half-mass) stellar radius in $\kpc$; (7)-(8) mean stellar $v/\sigma$ and stellar disc fraction, defined as in Section \ref{method}.}\\
\vspace*{15pt}
\end{tabular}
\end{table*}

\begin{table*}
\caption{Main properties for the sample of massive nearby spirals studied in this work.}
\label{tab:catalogue_obs} 
\centering
\begin{tabular}{lcccccccccc}
\hline\hline\noalign{\vspace{5pt}}
 Galaxy     &$\log_{10}(M_\star/M_\odot)$ & $\epsilon_{{M_\star},{\rm low}}$  & $\epsilon_{{M_\star},{\rm up}}$ & $\log_{10}(M_{\rm halo}/M_\odot)$ & $\epsilon_{{M_{\rm halo}},{\rm low}}$  & $\epsilon_{{M_{\rm halo}},{\rm up}}$ & $v_{\rm flat}$ & $\epsilon_{v_{\rm flat}}$ & $R_{\rm eff}$ & $\epsilon_{R_{\rm eff}}$  \\
      (1)&(2)&(3)&(4)&(5)&(6)&(7)&(8)&(9)&(10)&(11)\\
\noalign{\smallskip}
\hline\noalign{\vspace{5pt}}
NGC~7331 &         $10.78$&        $10.69$&        $10.84$&        $12.38$&        $12.21$&        $12.60$&       $239.00$&         $5.40$&         $3.99$&         $0.41$\\
     NGC~5985&        $10.91$&        $10.55$&        $11.10$&        $12.21$&        $12.12$&        $12.28$&       $293.60$&         $8.60$&        $10.71$&         $2.67$\\
    UGC~03205&        $10.94$&        $10.85$&        $11.00$&        $12.12$&        $11.95$&        $12.33$&       $219.60$&         $8.60$&         $5.35$&         $1.07$\\
    UGC~11914&        $10.95$&        $10.82$&        $11.04$&        $13.04$&        $12.44$&        $13.67$&       $288.10$&        $10.50$&         $3.12$&         $0.94$\\
    UGC~05253&        $10.95$&        $10.81$&        $11.05$&        $12.16$&        $12.08$&        $12.27$&       $213.70$&         $7.10$&         $4.28$&         $1.07$\\
     NGC~5907&        $10.96$&        $10.87$&        $11.01$&        $12.02$&        $11.93$&        $12.16$&       $215.00$&         $2.90$&         $7.88$&         $0.41$\\
     NGC~2998&        $10.98$&        $10.85$&        $11.07$&        $12.01$&        $11.91$&        $12.13$&       $209.90$&         $8.10$&         $7.06$&         $1.06$\\
     NGC~2841&        $11.00$&        $10.95$&        $11.04$&        $12.54$&        $12.42$&        $12.69$&       $284.80$&         $8.60$&         $5.51$&         $0.55$\\
     NGC~3992&        $11.01$&        $10.93$&        $11.07$&        $12.15$&        $12.03$&        $12.30$&       $241.00$&         $5.20$&         $9.99$&         $0.97$\\
    UGC~12506&        $11.12$&        $10.95$&        $11.19$&        $12.14$&        $11.96$&        $12.33$&       $234.00$&        $16.80$&        $12.36$&         $1.24$\\
     NGC~5371&        $11.13$&        $10.94$&        $11.26$&        $11.64$&        $11.53$&        $11.74$&       $209.50$&         $3.90$&         $9.80$&         $2.45$\\
    UGC~09133&        $11.15$&        $11.04$&        $11.24$&        $12.22$&        $12.18$&        $12.25$&       $226.80$&         $4.20$&         $5.92$&         $1.18$\\
     NGC~2955&        $11.17$&        $11.11$&        $11.22$&        $12.13$&        $11.80$&        $12.48$&       $  -\,^a$&         $-\,^a$&         $7.22$&         $0.72$\\
    UGC~02953&        $11.18$&        $11.03$&        $11.28$&        $12.29$&        $12.22$&        $12.36$&       $264.90$&         $6.00$&         $5.03$&         $1.51$\\
     NGC~6195&        $11.21$&        $11.15$&        $11.26$&        $12.16$&        $11.94$&        $12.42$&       $251.70$&         $9.30$&         $9.52$&         $0.95$\\
    UGC~11455&        $11.22$&        $11.11$&        $11.31$&        $12.61$&        $12.43$&        $12.84$&       $269.40$&         $7.40$&        $10.06$&         $1.51$\\
     NGC~0801&        $11.23$&        $11.18$&        $11.28$&        $12.00$&        $11.90$&        $12.14$&       $220.10$&         $6.20$&         $7.76$&         $0.78$\\
     NGC~6674&        $11.24$&        $11.15$&        $11.32$&        $12.42$&        $12.32$&        $12.56$&       $241.30$&         $4.90$&         $7.75$&         $1.54$\\
    UGC~02885&        $11.37$&        $11.30$&        $11.43$&        $12.62$&        $12.48$&        $12.79$&       $289.50$&        $12.00$&        $12.20$&         $1.22$\\
    UGC~02487&        $11.39$&        $11.33$&        $11.45$&        $12.58$&        $12.52$&        $12.67$&       $332.00$&         $3.50$&         $9.63$&         $1.45$\\
 ESO~563-G021&        $11.40$&        $11.33$&        $11.46$&        $12.93$&        $12.70$&        $13.21$&       $314.60$&        $11.70$&        $10.59$&         $1.59$\\
\noalign{\vspace{2pt}}\hline
\noalign{\vspace{5pt}}

\multicolumn{11}{p{1.0\textwidth}}{\textbf{Notes.} 
(1) Galaxy name; (2)-(4) stellar mass and related lower and upper uncertainties from \citetalias{Posti+19}; (5)-(7) halo mass and related lower and upper uncertainties from \citetalias{Posti+19}; (8)-(9) velocity of the flat part of the rotation curve (in $\kms$) and related uncertainty from \citet{Lelli+16}; (10)-(11) effective radius (in $\kpc$) and related uncertainty from \citet{Lelli+16}.}\\
\multicolumn{11}{p{1.0\textwidth}}{$^a$ The rotation curve of NGC~2955 does not have a well defined flat part, thus its $v_{\rm flat}$ is not reported in \citet{Lelli+16}.}\\
\vspace*{15pt}
\end{tabular}
\end{table*}

\begin{figure*}
\begin{center}
\includegraphics[width=0.88\textwidth]{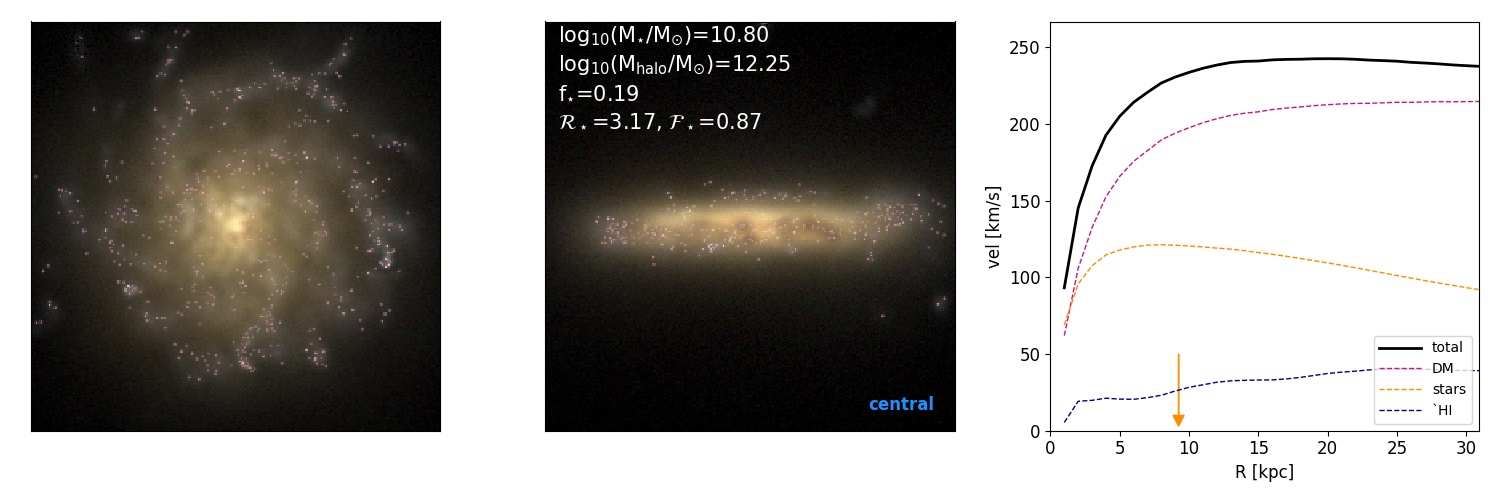}
\includegraphics[width=0.88\textwidth]{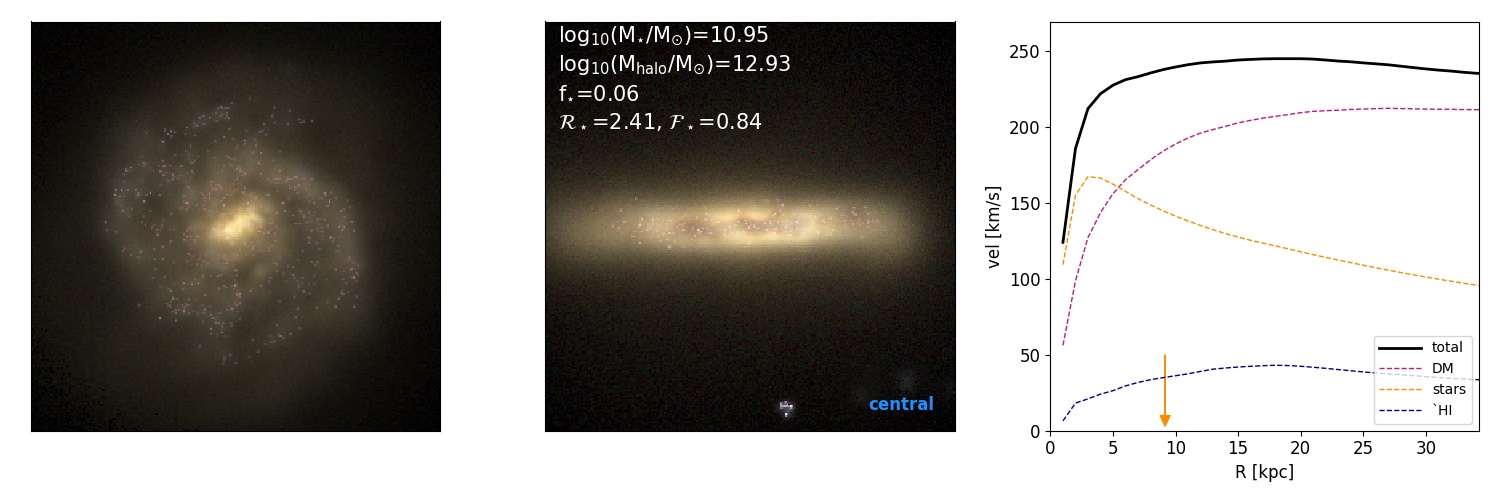}
\caption{Two examples of massive (central) disc galaxies from the EAGLE simulations (run Ref-L0100N1504). The left and central panels show the systems from a face- and edge-on perspective. The right panels show the total circular velocity profiles (solid black lines), along with the separate contributions (dashed lines) from stars (orange), gas (blue) and dark matter (magenta). The vertical arrow shows the half-M$_\star$ radius.} 
\label{fig:EAGLE_examples}
\end{center}
\end{figure*}

\begin{figure*}
\begin{center}
\includegraphics[width=0.88\textwidth]{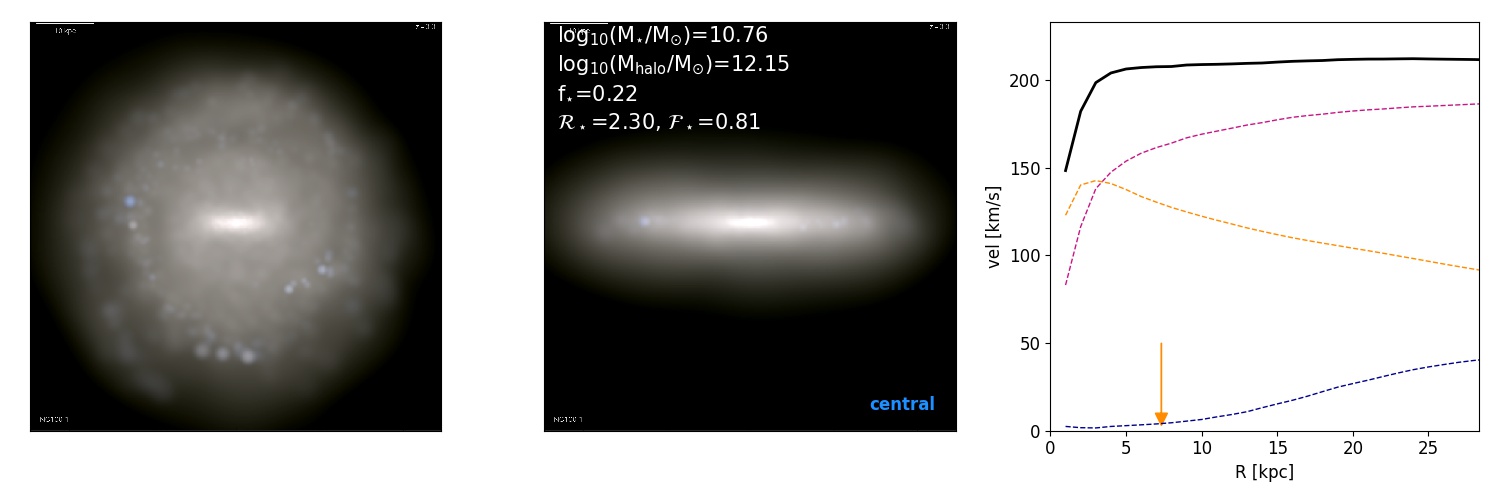}
\includegraphics[width=0.88\textwidth]{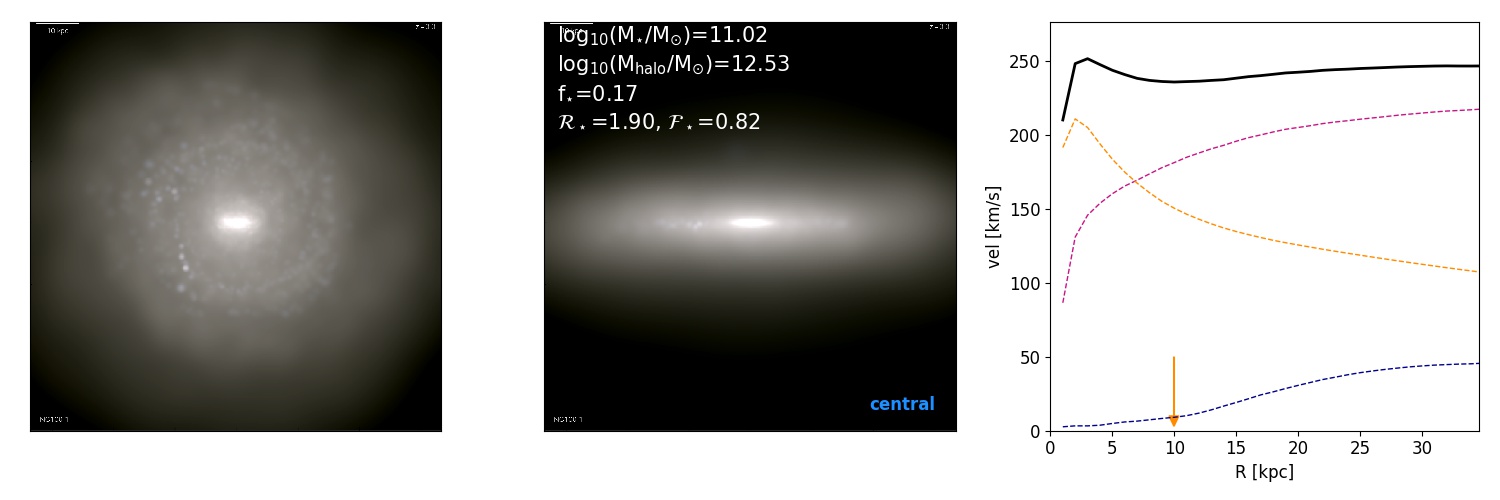}
\caption{As in Fig.\,\ref{fig:EAGLE_examples}, but for two massive (central) discs from the IllustrisTNG simulations (run TNG100-1).} 
\label{fig:TNG_examples}
\end{center}
\end{figure*}

In Fig.\,\ref{fig:EAGLE_examples} and \ref{fig:TNG_examples} we show face- and edge-on images for four representative massive disc galaxies extracted from the simulated sample studied in this work, along with their circular velocity profiles, truncated at the expected \hi\ radius, decomposed into the contributions from stars, gas and dark matter.
A full database of such figures for all simulated galaxies studied here can be freely downloaded at \url{https://drive.google.com/file/d/1WvnwRwAnOpEcGU9OTI-GHKf0w7-Ihlxn/view?usp=sharing}.

\section{Mass decomposition of synthetic rotation curves}\label{app:mass_dec}
\begin{figure*}
\begin{center}
\includegraphics[width=0.65\textwidth]{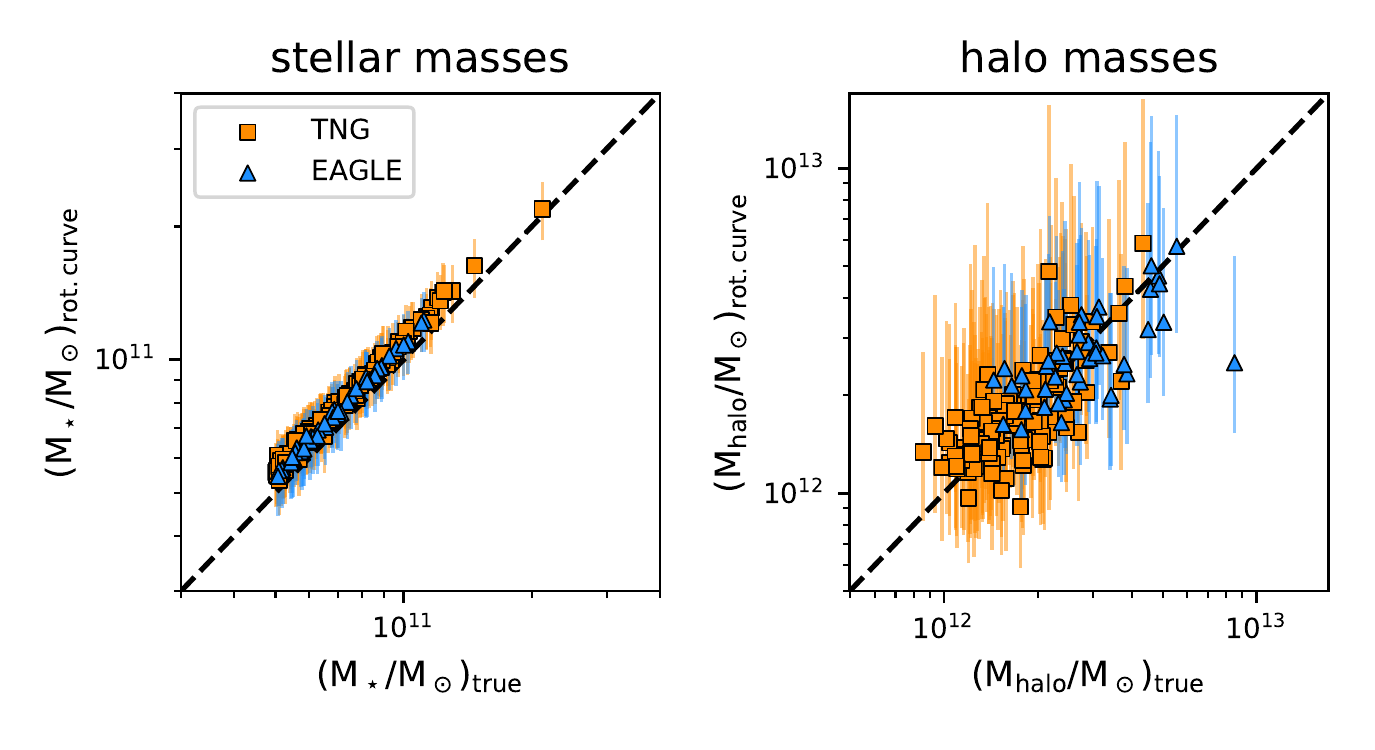}
\includegraphics[width=0.65\textwidth]{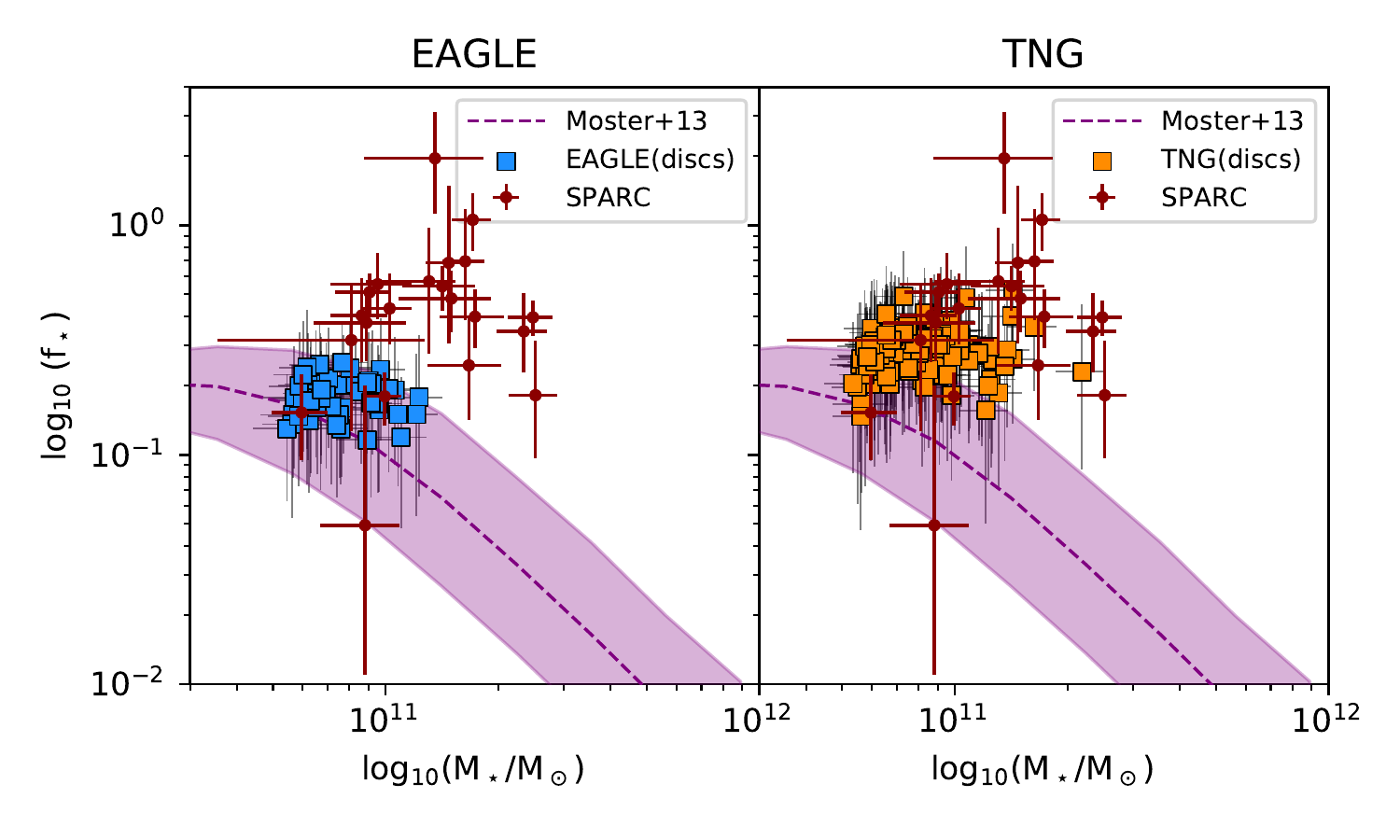}
\caption{\emph{Top-left panel}: comparison between the `true' stellar masses of the simulated disc galaxies ($x$-axis) and those derived via the decomposition of their synthetic rotation curve ($y$-axis). EAGLE (IllustrisTNG) galaxies are shown as blue triangles (orange squares). Error-bars are given from the difference between the 84th and 16th percentiles in the posterior probability distributions. The dashed line shows the one-to-one relation. \emph{Top-right panel}: the same, but for the halo masses. \emph{Bottom panels}: $f_\star-M_\star$ plot for the simulated discs using stellar and halo masses from rotation curve decomposition. Colors and symbols are as in Fig.\,\ref{fig:shmr}.} 
\label{fig:true_vs_observed}
\end{center}
\end{figure*}

We show here that our results do not change if we compute stellar and halo masses in the simulations using a procedure analogous to that of \citetalias{Posti+19}, based on the mass-decomposition of rotation curves.
To do so, we consider the synthetic rotation curves of our simulated galaxies truncated at their expected HI radius (see Section \ref{method}) and model them as 
\begin{equation}\label{vc}
    v_{\rm c}=\sqrt{v_{\rm DM}^2+ v_\star^2 + v_{\rm gas}^2}
\end{equation}
where $v_{\rm DM}$, $v_\star$ and $v_{\rm gas}$ are, respectively, the contributions of dark matter, stars and gas to the circular velocity profile.

As in \citetalias{Posti+19}, we assume Navarro-Frank-White \citep[NFW,][]{NFW} dark matter halo profiles. 
These are fully described by their virial mass $M_{200}$ and their concentration $c$, both of which are free parameters of the model. 
We assume $v_\star=\sqrt{\Upsilon\, v^2_{\star,{\rm true}}}$, where $v^2_{\star,{\rm true}}$ is given by $GM_\star(<R)/R$ and is assumed to be known, and $\Upsilon$ is a free parameter that mimics the effect of a (radially-constant) mass-to-light ratio.
With this parametrization, deviations from $\Upsilon=1$ correspond to variations in the inferred stellar mass with respect to its true value.
Finally, we fix $v^2_{\rm gas}$ to its true value ($GM_{\rm gas}(<R)/R$) as this gives only a minor contribution to the $v_{\rm c}$.
Following \citetalias{Posti+19}, the three free parameters of the model ($M_{200}$, $c$ and $\Upsilon$) are fit to the data via a Bayesian approach which adopts a prior on the $c-M_{200}$ relation motivated by N-body cosmological simulations \citep{DuttonMaccio14}.

In the top panels of Fig.\,\ref{fig:true_vs_observed} we compare the stellar and halo masses derived with this method with their `true' values taken from the simulation catalogues. 
Clearly, there is excellent agreement between true and inferred masses with consequently little variation in the $f_\star-M_\star$ relation (bottom panels of Fig.\,\ref{fig:true_vs_observed}).
We notice systematic shifts upwards for the inferred values of $f_\star$, with typical $\delta f_\star/f_\star$ of $22\%$ in EAGLE and $29\%$ in IllustrisTNG, which however fall well within the quoted uncertainties.
This is a confirmation of the validity of the \citetalias{Posti+19} method, and indicates that NFW halos, in the mass range considered here, are good proxies for the dark matter density profiles in EAGLE and IllustrisTNG.

\end{document}